\documentclass[pre,12pt,nofootinbib,twoside]{revtex4}
\usepackage{amsmath,amssymb,amsfonts,graphicx,bm}

\begin{document}

\def\as{\alpha_s}
\def\A{{\cal A}}
\def\B{{\cal B}}
\def\I{{\cal I}}
\def\gs{{\gamma^*}}
\def\ab{{\mathcal{A}\!\!\mathcal{B}}}
\def\GGqqg{|\Gamma^{(0)}_{\gamma^*\to q{\bar q}g}|^2}
\def\Gqq{\Gamma_{\gamma^*\to q{\bar q}}}
\def\bk{\bm k}
\def\bl{\bm \ell}
\def\br{\bm r}
\def\a{\alpha}
\def\b{\beta}
\def\al{\alpha_\ell}
\def\ae{\alpha_1}
\def\ake{{\bar\alpha}_1}
\def\az{\alpha_2}
\def\akz{{\bar\alpha}_2}
\def\ep{\epsilon}
\def\bLam{\bar \Lambda}
\def\blam{\bar{ \bm\lambda}}
\def\ka{\kappa}
\def\bs{\bar s_0}
\def\luv{\bm \ell\!-\!\mathrm{uv}}

${}$\hfill {\normalsize\raggedleft  DESY 04-116}\\[50pt]

\title{NLO Corrections to  the $\gamma^*$ Impact Factor:\\
First Numerical Results for the Real Corrections to  $\gamma_L^*$ }

\author{J.~Bartels$^{(a)}$ A.~Kyrieleis$^{(b)}$}

\affiliation{(a) II.~Institut f\"ur Theoretische Physik, Universit\"at
  Hamburg, \\
  Luruper Chaussee 149, 22761 Hamburg, Germany\\
  (b) Department of Physics \& Astronomy, University of Manchester, \\
  Oxford Road, Manchester M13 9PL, U.K.}

\begin{abstract}
We analytically perform the transverse momentum integrations in the real
corrections to the longitudinal $\gamma^*_L$ impact factor.
The resulting integrals are Feynman parameter integrals, 
and we provide a MATHEMATICA file which contains the integrands.     
The remaining integrals are carried out numerically. We perform a numerical 
test, and we compute those parts of the impact factor 
which depend upon the energy scale $s_0$:
they are found to be negative and, with decreasing values of $s_0$,
their absolute value increases.     
\end{abstract}

\maketitle


\section{Introduction}
The NLO corrections to the $\gs$ impact factor are calculated from
the photon-Reggeon vertices for $q\bar q$ and $q\bar qg$ production,
respectively. NLO corrections to the  $q\bar q$ intermediate state
involve the production vertex at one-loop level, $\Gqq^{(1)}$,
These virtual corrections have been calculated in
\cite{virt,kot}. As to the real corrections, the squared vertex
$\GGqqg$ is needed at tree level; it has been computed in \cite{real} and
\cite{combine} for longitudinal and transverse photon polarisation,
respectively (cf. also \cite{fad,kot}). In \cite{combine} we have combined the
infrared  divergences of the virtual and of the real parts,
and we have demonstrated their cancellation.
What remains to complete the NLO calculation of the photon impact factor
are the integrations over the $q\bar q$ and $q\bar qg$ phase space,
respectively. A slightly different approach of calculating the NLO
corrections of the photon impact factor has been proposed in
\cite{FIK}. Recently, the NLO calculation of another impact factor has
been completed \cite{IKP}, the impact factor for the transition
of a virtual photon to a light vector meson.

In this paper we perform, for the case of the longitudinal photon 
polarisation, the phase space integration in the real corrections. Our aim
is to have, as long as possible, analytic expressions which allow  
for further theoretical investigations. The main obstacle is the 
(infinite) integration over transverse momenta: in order to be able to perform 
the integration analytically, we introduce Feynman parameters.
This will allow for further theoretical investigations of the photon impact
factor. In particular, the Mellin transform of the real corrections
w.r.t the Reggeon momentum can be calculated. This representation
(together with an analogous representation of the virtual corrections) 
will also allow to study the impact
factor in the collinear limit and  to compare with known NLO results;
it can also  be a starting point for the resummation of the
next-to-leading logs(1/x) in the quark anomalous dimensions. 

Starting from the results of \cite{real},  
we have to integrate a sum of expressions, corresponding to 
products of Feynman diagrams \footnote{For simplicity we will 
in the following simply use `Feynman diagram' rather than `product of 
Feynman diagrams'.} that differ in their denominator structure. 
In order to introduce Feynman parameters we therefore  split this sum and
treat each Feynman diagram (or small groups of them) independently. This
gives rise to divergences, which in the sum of all diagrams will cancel
but show up in individual diagrams. 
The main task in the program of performing the integration analytically 
is the regularisation of these divergences in each individual diagram: 
for this we use the  subtraction method. 
After carrying out the integration over transverse momenta we
arrive at analytic expressions for each diagram. These
expressions are convergent integrals over the Feynman parameters and the 
momentum fractions of the quark and the gluon. Since the expressions 
are somewhat lengthy, we do not list them in this paper but 
provide a MATHEMATICA code which we describe in the appendix.   
As a first application of our results, 
we carry out the remaining parameter integrals numerically and perform
first tests of our calculation. In particular, we study the dependence
of the photon impact factor on the energy scale which is entirely contained
in the real corrections: it turns out to be in agreement with the 
expectations. 

Our paper is organized as follows. 
In the next section we will recall the structure of the real
corrections and specify the expressions we have to
integrate, following very much the strategy described in \cite{combine}. 
In the following section we introduce Feynman parameters 
and perform analytically the integration over transverse momenta.
The main emphasis lies on the regularisation of the 
divergences, which appear in the individual diagrams. 
Finally, the numerical results are discussed. 
An appendix describes the files that provide the Feynman 
parameter representations of the individual diagrams.    

\section{The real corrections}
\begin{figure}[htbp]
  \begin{center}
    \includegraphics[width=5cm]{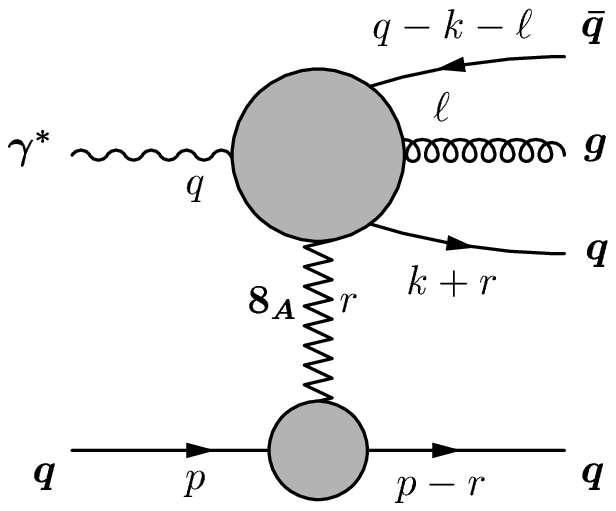}
    \caption{Kinematics of the process $\gamma^*+q' \to q\bar qg + q'$.
      \label{fig:qqgkin}}
  \end{center}
\end{figure}

The starting point for the real corrections  is 
the squared vertex $\GGqqg$ which, starting from the process 
$\gs+q\to q\bar qg+q$ for the longitudinal photon polarisation, 
has been calculated in \cite{real}. The notations are
shown in fig.\ref{fig:qqgkin}. The momenta are expressed in terms of
Sudakov variables: $k=\a q' + \beta p + k_\perp, \ell=\al \,q' +
\beta_\ell \,p + \ell_\perp$, with $q'=q+xp$, $Q^2=-q^2$ and $\bk^2=-k^2_\perp$.
\begin{figure}[htbp]
  \begin{center}
    \includegraphics[width=11.5cm]{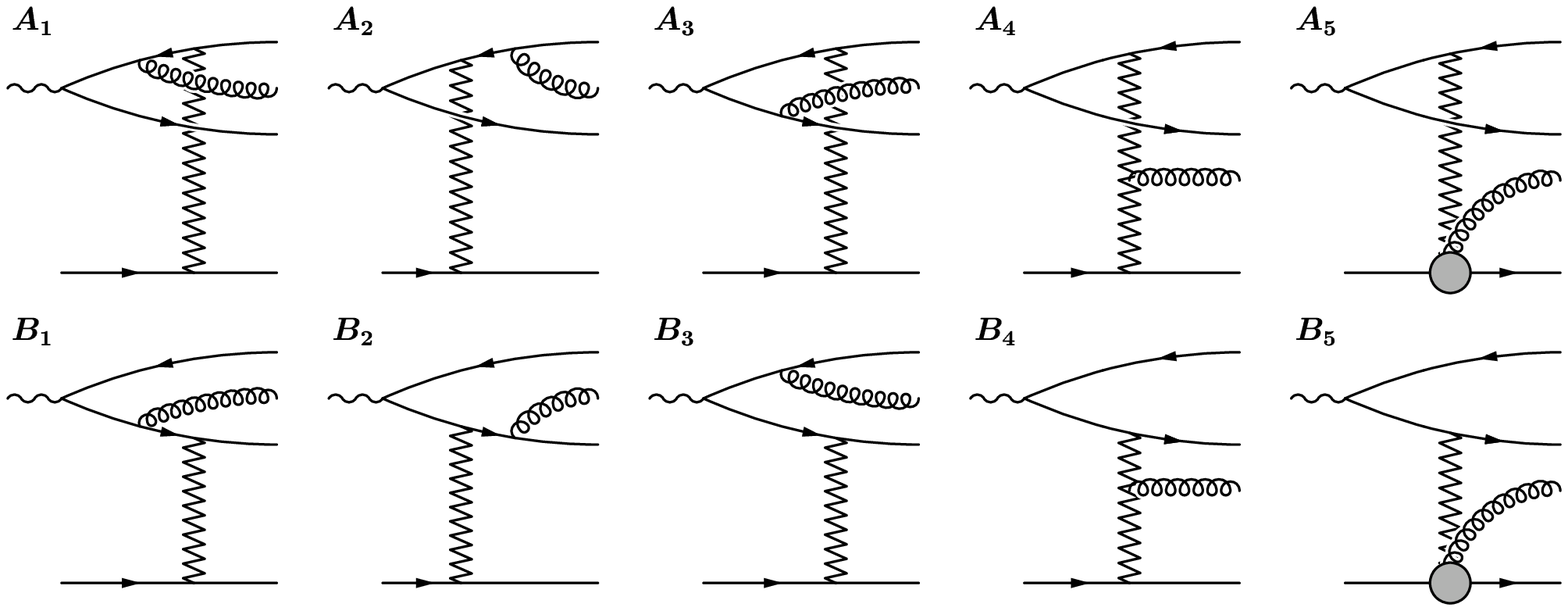}
    \caption{The diagrams contributing to $\Gamma_{\gs\to q\bar qg}$
      \label{fig:realdiags}}
  \end{center}
\end{figure}
Fig.\ref{fig:realdiags} shows the Feynman diagrams. The
product of two diagrams, summed over
helicities and colors of the external particles, is labelled following
the notation of the diagrams (e.g.:
$\A\B_{12}=A_1 B_2$). The expressions for the products $\A\A_{ij},
\A\B_{ij}, ...$ that we use here are taken from \cite{real}
where, with one exception, we have used the same notation: in the present 
paper we combine, for the sake of 
simplification, the diagrams 4 and 5 in the product with diagram 4, i.e. we set:
\begin{align*}
\A\A_{44}& + \A\A_{45} + \A\A_{54}\,\longrightarrow \A\A_{44}
,\nonumber\\
\A\A_{45} &, \A\A_{54} \,\longrightarrow 0,
\end{align*}
\begin{align*}
\A\B_{44}& + \A\B_{45} + \A\B_{54}\,\longrightarrow \A\B_{44}
,\nonumber\\
\A\B_{45} &, \A\B_{54} \,\longrightarrow 0,
\end{align*}
and similarly for $\B\A$ and $\B\B$. 
All products $\A\A$ etc. have (transverse) dimension $-6$; 
in order to deal with dimensionless expressions we multiply with 
$(Q^2)^3$. Let us introduce dimensionless variables:
\begin{align}
\label{norm}
\bk, \bl, \br, \Lambda, \sqrt{s_0} \,\rightarrow\, \frac{1}{|Q|}
(\bk, \bl, \br, \Lambda, \sqrt{s_0})
\end{align}
(the meaning of $\Lambda$ and $s_0$ will become clear in a
moment). From now on we will use only these new variables. 
We also define the abbreviations 
\begin{align*}
  \alpha_1 &\equiv \alpha\;,\\
  \bar \alpha_1 &\equiv (1-\alpha-\alpha_\ell)\;,\\
  \alpha_2 &\equiv (1-\alpha)\;,\\
  \bar \alpha_2 &\equiv (\alpha+\alpha_\ell).
\end{align*} 
Furthermore, we introduce the label $\ab$, in order to denote a generic product
of amplitudes ($\A\A, \A\B, \B\A$ or $\B\B$):
\begin{align}
\label{eq:def_ab}
\ab &=\frac{(Q^2)^3}{\alpha_1 \bar \alpha_1} (\A\A_{11}\;\mbox{or}\;
\A\A_{12}\;\mbox{or} \;\A\A_{21}\, \ldots\,
\mbox{or} \;\A\B_{11}.
\ldots).
\end{align}
It is then easy to see that these $\ab$'s are dimensionless and only depend on
the dimensionless momenta $\bk, \bl$ and $ \br$.
In (\ref{eq:def_ab}) we have included a part of the $q\bar qg$ phase
space measure in the definition of the $\ab$'s, in order to simplify the 
expressions below. Finally, $\GGqqg$ will be proportional to the sum of all 
the $\ab$.

The procedure of arriving at finite NLO corrections to the $\gs$ impact
factor has been described in \cite{combine}. Let us briefly review the main
steps. $\GGqqg$ has to be integrated over the $q\bar qg$ phase space. 
Before doing this integration, two restrictions have to be observed.
First, we have to exclude that region of phase space 
where the gluon is separated in rapidity from the $q\bar q$ pair (central
region); this configuration belongs to the LLA and has to be subtracted. 
To divide the $q\bar qg$ phase space an energy cutoff $s_0$ is introduced. 
This energy cutoff plays the role of the energy scale: when combining 
the NLO impact factor with the NLO BFKL Green function it will be important  
to use the same scale in all pieces. Since the virtual corrections to the 
NLO impact factor are independent of $s_0$, all dependence on the energy 
scale resides in the real corrections. As a result, the calculations 
described in this paper can already be used to study the $s_0$ dependence 
of the NLO impact factor. Next, we need to take care of the infrared
infinities. The divergences in $\GGqqg$ due to the gluon
being either soft or collinear to either of the fermions are
regularised by subtracting the approximation of the squared vertex in
the corresponding limit. These expressions are then re-added and integrated
in $D=4-2\ep$ space-time dimensions, giving rise to poles in $\ep$,
which drop out in combination with the virtual corrections and to 
finite pieces. The subtraction of the collinear limit requires 
the introduction of a momentum cutoff parameter, $\Lambda$. The final  
NLO corrections must be independent of this auxiliary parameter.
In our numerical analysis we will perform this important test.  

According to \cite{combine} the full NLO corrections to the $\gs$ impact 
factor have the following form:
\begin{align}
\label{eq:phi1}
  \Phi_{\gs}^{(1)} = &
  \left.\Phi_{\gs}^{(1,{\mathrm virtual})}\right|^{\mathrm finite}
  - \frac{2 \Phi_{\gs}^{(0)}}{(4\pi)^2} \left\{
    \beta_0 \ln\frac{\br^2}{\mu^2} + C_F\ln(\br^2) 
  \right\}
\nonumber\\
&+\int_0^1 d \alpha\,\int\,\frac{d^{2-2\ep}\bk}{(4\pi)^2}\,{\mathcal{I}}_2(\alpha, \bk)\,\bigg\{
  C_A \left[\ln^2\alpha(1-\alpha)s_0 -\ln^2 M^2\right]\nonumber\\
  &\qquad\qquad + 2C_F \left[8 - 3\ln \alpha(1-\alpha) \Lambda^2 + \ln^2 M^2
  + \ln^2 \frac{\alpha}{1-\alpha} \right]
\bigg\}\nonumber\\
&+ C_A \left.\Phi_{\gs}^{(1,{\mathrm real})}\right|_{C_A}^{\mathrm finite}
+ C_F \left.\Phi_{\gs}^{(1,{\mathrm real})}\right|_{C_F}^{\mathrm finite}\;.
\end{align}
The terms multiplying $\Phi_\gs^{(0)}$ stem from the UV renormalisation 
which belongs to the virtual corrections and does not need to be 
discussed in the present context. 
The finite pieces which are left after combining the IR singular pieces 
of the virtual and real corrections above are  given in the second and 
third lines of eq.\eqref{eq:phi1}. ${\mathcal{I}}_2$ is proportional 
to the squared LO photon wave function:
\begin{equation}
\label{eq:bornifl}
\I_{2} (\alpha,\bk) =\frac{2e^2
 e_f^2\sqrt{N_c^2-1}}{(2\pi)^{3-2\epsilon} (Q^2)^{2\ep}} 
 \alpha^2 (1-\alpha)^2 
 \left(\frac{1}{D(\bk)} - \frac{1}{D(\bk+\br)}\right)^2
\end{equation}
with $D(\bk) = \bk^2 + \alpha(1-\alpha)$. Finally, the squared mass 
of the quark-antiquark pair, $M^2$, is a function of $\alpha, \bk ,\br$, 
and it is not important for our present analysis. 

Let us now focus on the last line of eq.\eqref{eq:phi1}. In terms of the
products of single diagrams, $\ab$, these finite contributions to
the real corrections \cite{combine} read:
\begin{align}\label{eq:sum1CF}
 &\left.  C_F\: \Phi_{\gs}^{(1,{\mathrm real})}\right|_{C_F}^{\mathrm
 finite} = \frac{e^2 e_f^2 \sqrt{N_c^2-1} }{ (2 \pi)^{6 - 4\ep} (Q^2)^{2\ep}}  
 \int d^{2-2\ep}\bk\, d^{2-2\ep}\bl \int_0^1 d\alpha\int_0^{1-\alpha}
 \frac{d\al}{\al}\, \nonumber\\
&\hspace{.7cm}\times\sum_{C_F} 
 \Biggl\{ \ab - \ab_{\mathrm{soft}} - \left( \ab_{
 \bar{\mathrm{q}} }- \ab_{\bar{\mathrm{q}},\,soft} \right)
 \Theta(\alpha_\ell \Lambda - |\bl '|) \nonumber\\
&{}\hspace{1.8cm}- \left( \ab_{\mathrm{q}} - \ab_{{\mathrm{q}},\,soft}
 \right) \Theta(\alpha_\ell \Lambda - |\bl\, ''|)
\Bigg\}, 
\end{align}
\begin{align}
\label{eq:sum1CA}
 \left.  C_A \:\Phi_{\gs}^{(1,{\mathrm real})}\right|_{C_A}^{\mathrm
 finite}&= \frac{e^2 e_f^2 \sqrt{N_c^2-1} }{ (2 \pi)^{6-4\ep}(Q^2)^{2\ep} } 
 \int d^{2-2\ep}\bk\, d^{2-2\ep}\bl  \int_0^1 d\alpha \Bigg\{ \int_0^{1-\alpha}
 \frac{d\alpha_\ell}{\al} \nonumber \\
&\hspace{-.7cm}\times\sum_{C_A}  \bigg[ 
 \ab  -  \ab_{\mathrm{soft}} - \left(
 \ab_{\mathrm{cen}}-\ab_{\mathrm{cen,\,soft}} \right)
 \Theta(|\bl|-\alpha_\ell \sqrt{s_0}) 
  \bigg]\nonumber \\
-& \int_{1-\alpha}^{\infty}
 \frac{d\al}{\al}\,
 \sum_{C_A} \ab_{\mathrm{cen}}\,\Theta(|\bl|-\alpha_\ell \sqrt{s_0}) \Bigg\}.
\end{align}
The evaluation of these two expressions is the issue of this
paper. The sums extend over the $C_F$ and $C_A$ parts of the $\ab$'s,
respectively. Note that the sum over all $\ab$'s is essentially $\GGqqg$. 
Whereas the full sum is finite, the individual contributions 
$\ab$'s, without further modifications, would be divergent: our task, 
therefore is to render them finite, diagram by diagram. This is the content 
of eqs.(\ref{eq:sum1CF}) and (\ref{eq:sum1CA}). 
In order to indicate the different limits where
the integral of $\ab$ diverges we use the following subscripts: 
\begin{align} 
\label{eq:limits}
\mathrm{soft} &:=\; \alpha_\ell \sim |\bm \ell| \to 0\nonumber\\
\mathrm{\bar q} &:=\;  |\bm \ell'| \to 0\nonumber\\
\mathrm{ q} &:=\;  |\bm \ell''| \to 0\nonumber\\
\mathrm{cen} &:=\;  \alpha_\ell \to 0 \nonumber\\
\ab_\mathrm{x, \,soft} &:= \left[ \ab_\mathrm{x}
\right]_\mathrm{soft}.
\end{align}
Here $\bl'$ \and $\bl\,''$ denote the two collinear vectors
\begin{align}
\label{eq:collvecs}
\bl' = \bl + \frac{\alpha_\ell}{\az} \bk
\quad,\quad\bl\,'' = \bl
- \frac{\alpha_\ell}{\ae} \bm (\bk+\br) .
\end{align}  
The parameter $\Lambda$ defines a cone around the collinear directions,
specifying thereby the region of the subtraction. 
The exclusion of the central configuration (cen) is realised by
subtraction: only the rapidity region $\alpha_\ell > |\bl|/\sqrt{s_0}$ 
is counted as a contribution to the impact factor.


\section{The Integration}

Our aim is, in eqs.(\ref{eq:sum1CF},\ref{eq:sum1CA}), the analytic 
integration over $\bk$ and $\bl$. 
To this end we introduce Feynman parameters. Since different  
Feynman diagrams provide different denominators, we have to deal 
the $\ab$'s independently rather than in the sum
of all (as an alternative, an attempt the find common denominators 
would lead to expressions that have too lengthy numerators). 
As indicated in eqs.(\ref{eq:sum1CF},\ref{eq:sum1CA}),
in each single diagram $\ab$ we have to combine
the full expression with subtractions which are obtained from 
certain approximations (soft, cen,...), given by
eq.\eqref{eq:limits}. As a first step, we analyse the divergences   
of all diagrams. The ta\-ble \ref{tab:ABdiv} shows, 
which diagram diverges in which limit. 
Both in the left hand and in the right hand part of the table, it is 
the second column which lists the divergent limits which 
in eqs.(\ref{eq:sum1CF},\ref{eq:sum1CA}) require a suitable subtraction. 
The third columns contain an additional divergence that appears due to the 
separate treatment of the diagrams; we shall take care of them by 
making a further appropriate subtraction form each $\ab$. 
\begin{table*}
\begin{tabular}{||l c c | c c c c | c c c }
 $\A\A$  &  &  &$\:$ cen & cen/soft & soft & coll$\:$  &$\:$ l-uv & is & lr\\
\hline
11 & $C_F$ & & & & & & $\times$\\
12 & $C_F$,&  $C_A$ & & & & $\bar q$ & $\times$\\ 
22 & $C_F$ & & & & &  $\bar q$ & $\times$\\[.2cm]
13 & $C_F$ & & & & & $q$ \\ 
23 & $C_F$,& $ C_A$ & & & $\times$ & $q,\,\bar q$\\
33 & $C_F$ &  & & & & $q$ \\ 
34 & & $C_A$ & & &  & $q$ \\[.2cm]
15 & & $C_A$ & $\times$ & $\times$ & & & & $\times$ & $\times$\\  
25 & & $C_A$ & $\times$ & $\times$ & $\times$ & $\bar q$& & &$\times$ \\ 
35 & & $C_A$ &  &  & $\times$  & $q$ & & $\times$\\[.2cm]
14 & & $C_A$ & $\times$ & & &  & $\times$ & & $\times$\\  
24 & & $C_A$ & $\times$ & & & $\bar q$ & $\times$ & & $\times$\\
44 & & $C_A$ & $\times$ & & & & $\times$ & & $\times$ \\
\end{tabular}
\begin{tabular}
{||l c c | c c c c | c  c c }
 $\A\B$  &  &  & $\:$cen & cen/soft & soft & coll$\:$ &$\:$ l-uv &
is & lr \\
\hline
11 & $C_F$, & $C_A$ & & & & & \\
12 & $C_F$& & & & & $ q$ & \\ 
22 & $C_F$, &$C_A$ & & & $\times$ & $q,\, \bar q$ & \\[.2cm]
13 & $C_F$, &$C_A$ & & & & $\bar q$ \\ 
23 & $C_F$& & & & & $\bar q$ \\
33 & $C_F$, &$C_A$  & & &  $\times$  & $q,\,\bar q$ \\ 
34 & & $C_A$ & & &  & $q$ \\[.2cm]
15 & & $C_A$ & $\times$ & $\times$ & & &  &$\times$& $\times$ \\  
25 & & $C_A$ & $\times$ & $\times$ & $\times$ & $\bar q$& & &$\times$ \\ 
35 & & $C_A$ &  &  & $\times$  & $q$ & &  $\times$ \\[.2cm]
14 & & $C_A$ & $\times$ & & &  & & & $\times$\\  
24 & & $C_A$ & $\times$ & & & $\bar q$ &  &  &$\times$\\
44 & & $C_A$ & $\times$ & & & & &  &$\times$      
\end{tabular}
${}\vspace{.4cm}$
\caption{The divergences of the diagrams $\A\A_{ij}$ and $\A\B_{ij}$ \label{tab:ABdiv}}
\end{table*}
Let us go through these subtractions in some detail.
Each product $\ab$ has the general structure
\begin{equation}
\label{eq:integform}
\ab = \frac{Z}{D_i D_j D_k D_l}\quad ,\qquad D_n = D_n (\bk, \bl, \br).
\end{equation}
The denominators $D_n$ are given by (here we use definitions which 
slightly differ from those of \cite{real}):
\begin{align}
D_1 &= \,\alpha_\ell \left[ \alpha_1 {\bar
    \alpha}_1  + \alpha_1 (\bk + \bl + \br)^2 + {\bar \alpha}_1
    (\bk + \br)^2 \right] + \alpha_1 {\bar \alpha}_1 \bl^2
        \label{eq:D1} \\
D_2 &= \,\alpha_1 \alpha_2  + (\bk + \br)^2 \\
D_3 &=  (\alpha_1\, \bm\ell -\alpha_\ell \,
(\bm k + \bm r)  )^2 \,=\ae^2\, \bl^{''2}\\
D_4 &=( \bar{\alpha}_1 \, \bm\ell + \alpha_\ell \, (\bm k + \bm\ell)
)^2 \,=\az^2\, \bl^{'2}\\
D_5 &=  \,\alpha_\ell \left( \alpha_1 {\bar
    \alpha}_1  + \alpha_1 (\bk + \bl)^2 + {\bar \alpha}_1
    (\bk + \br)^2 \right) + \alpha_1 {\bar \alpha}_1 (\bl - \br)^2\\
D_6 &= \al \left[ \ae \ake \,   + \ake \bk^2 + \alpha_1
\, (\bm k + \bm\ell)^2 \right] +  \alpha_1 \ake \, \bm\ell^2\\
D_7 &= \ake \akz \,  + (\bm k + \bm\ell)^2\; \label{eq:D7}.
\end{align}
Note that, in some of the  $\ab$'s, two of the four $D's$ 
in eq.\eqref{eq:integform} may coincide. The
numerator $Z$ is a polynomial in scalar products of the (dimensionless) 
transverse momenta:
\begin{align}
\label{eq:Z}
Z = \lambda_1\, \bk^2 + \lambda_2\, \bl^2 +  \lambda_3\,
\bk\bl  + \lambda_4\, \bk\br  + \lambda_5\, \bl\br  +
\lambda_6\, \br^2 + \lambda_7\:,\quad \lambda_i=\lambda_i(\a, \al).
\end{align}
(without our choice of dimensionless momenta, (\ref{norm}), 
the last term, $\lambda_7$, would have been proportional to $Q^2$).  
Let us now consider one single $\ab$. Using
\begin{align*}
\prod_{i=1} \frac{1}{D_i^{p_i}} = \frac{\Gamma(\sum_i
    p_i)}{\prod_i\Gamma(p_i)}\;\int_0^1 \prod_i\left(d\b_i
    \,\b_i^{p_i-1}\right) \,\frac{\delta(1-\sum_i \b_i)}{\left[\sum_i
    \b_i D_i \right]^{\sum_ip_i} }.
\end{align*} 
we introduce the Feynman parameter representation:
\begin{align}
\label{eq:afterb}
\ab =  6  \int_0^1 \;&\prod_i d\b_i \;\delta(1-\sum \b_i)\;f_\b \:\:
\frac{Z}{\left[\sum \b_i D_i\right]^4}.
\end{align}
The $\b_i$ carry the same labels as the denominators $D_i$ they belong
to, and the sum in the denominator of (\ref{eq:afterb}) is understood 
to extend over those indices that occur in the diagram. $f_\b$ stands
for products and powers of the $\b's$ in the numerator which appear in case 
of some of the four $D's$ being equal. For example, 
the product $\A\B_{13}$ reads
\begin{equation}
\frac{Z}{D_1^2\, D_2 \,D_3 } = 6  \int_0^1 d\b_1\,d\b_2
  \,d\b_3\;\frac{\delta(1-\sum
  \b_i)\:\b_1\;Z}{\left[\b_1 D_1 + \b_2 D_2 + \b_3 D_3\right]^4}.
\end{equation}

In order to determine  the Feynman representation of the approximations of $\ab$, we
first consider the numerator and the denominator, in
eq.(\ref{eq:integform}), in the appropriate approximation; we denote  
 the result by a subscript. Then, the  Feynman
representation is introduced as just described:  
\begin{align}
\label{eq:xafterb}
\ab_x =  6  \int_0^1 \;&\prod_i d\b_i \;\delta(1-\mbox{$\sum$} \b_i)\:f_\b \:\:
\frac{Z|_x}{\left[\mbox{$\sum$} \b_i \left. D_i\right|_x\right]^4}\:,\quad
x=\mbox{$q$, $\bar q$, cen, ...}
\end{align}
Each limit in the original momentum space representation 
(in the $\{\bk, \bl, \al\}$ space) 
unambiguously corresponds to a `corner' in the $\{\bk, \bl, \al, \b_i\}$ space 
of the Feynman parameter representation. By our way of calculating the
Feynman representation of $\ab_\mathrm{soft}, \ab_\mathrm{q},
\ab_\mathrm{\bar q}, \ldots$, \eqref{eq:xafterb}, we 
ensure that the cancellation between $\ab$ and its approximations,
which originally was formulated in the momentum space Feynman amplitudes,
remains valid also after the introduction of Feynman parameters. The
prescription \eqref{eq:xafterb} in particular ensures that our
original expressions for the real corrections, eqs.(\ref{eq:sum1CF},
\ref{eq:sum1CA}),  are exactly
translated into the Feynman parameter space.

We now perform the integrations over $\bk$ and $\bl$.
According to the different integration regions that occur in
eqs.(\ref{eq:sum1CF},\ref{eq:sum1CA}) we are faced with three different 
types of integrals. Let us discuss them in some detail. 
For convenience, we generalize to $2-2\epsilon$ transverse dimensions.   


$\bullet\quad\int\:\ab$

Let us consider the integration $\int d^{2-2\ep}
\bk\;d^{2-2\ep} \bl \hspace{.1cm} \ab$. The $\bl$-integration extends
over the whole space. After the introduction of Feynman parameters,
eq.(\ref{eq:afterb}), we have
to integrate
\begin{align}
\label{eq:fulltodo}
J_1 := \int d^{2-2\ep} \bk\;d^{2-2\ep} \bl\:
\frac{Z}{\left[\sum \b_i D_i\right]^4}.
 \end{align} 
The generic structure of transverse momenta is:
\begin{align}
\label{eq:I1_1}
J_1 = \int d^{2-2\ep} \bk\;d^{2-2\ep} \bl\:
\frac{\lambda_1\, \bk^2 + \lambda_2\, \bl^2 +  \lambda_3\,
\bk\bl  + \lambda_4\, \bk\br  + \lambda_5\, \bl\br  +
\lambda_6\, \br^2 + \lambda_7}{\left[\tau \,\bk^2 + \rho \,\bl^2 + 2\, \gamma
\,\bk\bl + 2 \eta \,\bk\br + 2 \mu \,\bl\br + D_r \right]^4}.
 \end{align} 
The coefficients $\lambda_i$ in the numerator depend upon $\a, \al$.
The coefficients in the denominator, $\tau, \rho, \gamma, ...$, 
are functions of $\a, \al$ and of the Feynman parameters $\b_i$. 
$D_r$ depends, in addition, on $\br^2$ 
(and on $Q^2$, which has been normalized to $1$).
In order to eliminate, in the denominator, the mixed scalar products, 
we perform the shifts:
\begin{align}
\label{eq:fushi}
\bk &= \bk' - \frac{\gamma}{\tau} \bl' - \frac{1}{\mbox{det} \Omega} (\rho \eta - \gamma \mu)
\br \nonumber\\
\bl &= \bl' -  \frac{1}{\mbox{det} \Omega} (\tau \mu - \gamma
\eta) \br.
\end{align}
We arrive at:
\begin{align}
\label{eq:I1_2}
J_1 =& \int d^{2-2\ep} \bk'\;d^{2-2\ep} \bl'\; \frac{A\, \bk^{'2} + B\,
\bl^{'2} + C }{[\tau \bk^{'2} +\omega \bl^{'2} + R]^4},
\end{align}
where we have introduced the definitions
\begin{equation}
\label{eq:omROm} 
\omega = \frac{1}{\tau}\,\mbox{det} \Omega \:,\quad R = D_r -
\frac{1}{\mbox{det} \Omega } ( \eta^2 \rho +  \tau\mu^2 - \eta \mu \gamma) \br^2 \:,\quad \Omega = \left( \begin{array}{cc}
\tau & \gamma \\
\gamma & \rho 
\end{array} \right).
\end{equation}
In the numerator of (\ref{eq:I1_2}) we have already dropped the mixed scalar 
products $(\bk'\,\bl',\bk'\,\br', \bl'\,\br')$: when performing, in 
(\ref{eq:I1_1}), the shifts, such scalar products appear. However, 
after angular integration their contribution vanishes. Note that except for
$A=\lambda_1$ all coefficient functions appearing in eq.\eqref{eq:I1_2}, 
in general, depend on the  Feynman parameters $\beta_i$ and on $\a, \al$. 
Performing the momentum integrations we get:
\begin{align}
\label{eq:full}
J_1 =& \int d^{2-2\ep} \bk\;d^{2-2\ep} \bl\; \frac{A\, \bk^2 + B\,
\bl^2 + C }{[\tau \bk^2 +\omega \bl^2 + R]^4} \nonumber\\
=& \, \frac{\pi^{2-2\ep}}{6\,(\tau \omega)^{1-\ep} R^{1+2\ep}}
\,\Bigg\{ (1-\ep)\Gamma(1-2\ep)\,\left( \frac{A}{\tau } \; + \;
\frac{B}{\omega} \right) \, +\Gamma(2+2\ep) \frac{C}{R}\, \Bigg\}.
\end{align}
Note that the integral converges for $\ep=0$. We will need the general
result ($\ep \ne 0$) later.
As our final result for the Feynman
parameter representation of $\ab$ we define a quantity $X$ by 
\begin{align}
\label{eq:X}
\int d^2 \bk\;d^2 \bl \:\ab &=  6  \int_0^1 \;\prod_i
d\b_i \;\delta(1-\mbox{$\sum \b_i$})\:\, f_\b
\:\left.J_1\right|_{\ep=0} \nonumber\\
&= \int \prod_i d\b_i\;\delta(1-\mbox{$\sum \b_i$})\:\; X.
\end{align}
For each diagram $\ab$, the function $X$ is given as MATHEMATICA code, 
described in the appendix.   

In order to integrate the soft ($\bl \sim \al \to 0$) approximation of
$\ab$ we  consider the numerator
and the denominator of $\ab$ in the soft approximation and  introduce
the Feynman parameter representation
according to  \eqref{eq:xafterb}. We then  carry out the momentum
integration exactly as just described ending up with an expression 
which we call $X_\mathrm{soft}$:
\begin{align}
\label{eq:xsoftafterb}
\int d^2 \bk\;d^2 \bl \:\ab_\mathrm{soft} 
&= 6  \int \;\prod_i d\b_i \;\delta(1-\mbox{$\sum$} \b_i)\:
\int d^2 \bk\;d^2 \bl \frac{f_\b \:Z|_\mathrm{soft}}{\left[\mbox{$\sum$} \b_i
 \left. D_i\right|_\mathrm{soft}\right]^4}\nonumber\\ 
&=  \int \prod_i d\b_i\;\delta(1-\mbox{$\sum \b_i$})\:\; X_\mathrm{soft}.
\end{align}

It is instructive to see how the soft limit      in
the momentum space translates into the Feynman parameter space.
The corresponding region in the Feynman parameters space is 
determined by the behaviour of the denominators $D_i$ in the soft limit.  
One obtains
$\sum\b_i D_i|_\mathrm{soft}$ from $\sum\b_iD_i$  by `weighting'
 $\{\b_2, \b_5, \b_7\} \to \{\b_2, \b_5, \b_7\}  
\rho^2$, $\{\b_1, \b_6, \bl, \al\} \to \{\b_1, \b_6, \bl, \al\} \rho$,
and then expanding around $\rho= 0$ and by keeping only the most divergent
term. The prescription
for the $\b$'s remains unchanged after the integration over $\bl$
and $\bk$; the soft region is therefore specified by:
\begin{align}
\label{eq:softlimit}
\mathrm{soft:}\qquad \b_2, \b_5, \b_7 \quad \ll \quad \b_1, \b_6, \al
\quad \ll \quad \b_3, \b_4.
\end{align}
Hence, we can obtain $X_\mathrm{soft}$ either in the way given in
eq.\eqref{eq:xsoftafterb} or by considering $X$ (eq.\eqref{eq:X}) in the limit
\eqref{eq:softlimit} (indicated by the subscript):
\begin{align}
\int &d^2 \bk\;d^2 \bl \:\ab_\mathrm{soft} \nonumber \\
&= 6  \int \;\prod_i d\b_i \;\delta(1-\mbox{$\sum$}\b_i)\:
\int d^2 \bk\;d^2 \bl \left[\frac{f_\b\;Z}{\left[\mbox{$\sum$}\b_i
D_i\right]^4}\right]_{\textstyle  \b_2, \b_5, \b_7 \ll \b_1,\b_6,\al,\bl \ll
\b_3, \b_4} \nonumber\\ 
&=  \int \prod_i d\b_i\;\delta(1-\mbox{$\sum$}\b_i)\:\;
\left[X\right]_{\eqref{eq:softlimit} } \label{eq:softafterb}
\end{align}
Note that we must not apply  the approximation
\eqref{eq:softlimit} to the argument 
$\delta$-function in eq.\eqref{eq:softafterb} (otherwise, the translation of the real
corrections from the momentum space to the Feynman parameter space
would be incorrect). This has the following consequence. After
choosing any  of the $\b$-integrations to be
done by means of the $\delta$-function we have
\begin{align}
\label{eq:deltastep}
\int d\b_k \prod_i d\b_i\;\delta(1-\mbox{$\sum \b_i$})\:\left( X -
X_\mathrm{soft}\right) = \int \prod_i d\b_i\;\:\left( \tilde{X} -
\tilde{X}_\mathrm{soft}\right)
\end{align}
with the  tilde symbol on the rhs indicating that $\b_k$ is expressed in terms of the
other $\b_i$. Since the argument of the $\delta$-function is not
approximated,
$\tilde{X}_\mathrm{soft}$ is not equal to $\tilde{X}$ taken in the limit
\eqref{eq:softlimit}. However, both $\tilde{X}_\mathrm{soft}$ and
$\tilde{X}$ become equal in this limit. 


$\bullet\quad\int^\Lambda \ab_\mathrm{coll}$

Next we turn to the collinear approximation of $\ab$. 
We use the label 'coll' to denote the two collinear
limits  which are defined as either $|\bl''|\to 0$ or $|\bl'|\to 0$ 
(cf. eqs.(\ref{eq:limits}) and (\ref{eq:collvecs})). Following the
notation introduced in eq.\eqref{eq:xafterb} we use the  collinear approximation of the numerator of $\ab$ and of the $D's$.
The region of integration is restricted to a cone around the collinear 
direction:
\begin{align}
\label{eq:colltodo}
J_2 := \int d^{2-2\ep} \bk\;d^{2-2\ep} \bl\:\:\Theta(\al \Lambda -
|\bl_c|)\:\:
\frac{Z_\mathrm{coll}}{\left[\sum \b_i
\left.D_i\right|_\mathrm{coll}\right]^4} \:,\:\quad\bl_c = \bl', \bl''. 
 \end{align} 
 Starting from eq.\eqref{eq:colltodo}
we first express $\bl$ in
eq.\eqref{eq:colltodo} through $\bl_c$. The list of the $D's$ in 
eqs.(\ref{eq:D1}) - (\ref{eq:D7}) 
shows that in the limit $\bl'\to 0$, in leading order, all $D_i$ become 
independent of $\bl'$, except for $D_4$ which is proportional to
$\bl^{'2}$.  The same holds for the the second collinear
limit, with $D_3$ instead of $D_4$. The denominator in
eq.\eqref{eq:colltodo} therefore depends on $\bl_c$ only via $\bl_c^2$
and the integrand is parametrised as 
\begin{align}
\label{eq:I2_1}
J_2 = \int d^{2-2\ep} \bk\;& d^{2-2\ep} \bl_c\:\Theta(\al\Lambda-|\bl_c|)\nonumber\\
&\times \frac{\lambda_1\, \bk^2 + \lambda_2\, \bl_c^2 +  \lambda_3\,
\bk\bl_c  + \lambda_4\, \bk\br  + \lambda_5\, \bl_c\br  +
\lambda_6\, \br^2 + \lambda_7}{\left[ \tau \,\bk^2 + \rho \,\bl_c^2 + 2 \eta \,\bk\br  + D_r \right]^4}.
 \end{align} 
Note that, since we are using $\bl_c$ instead of $\bl$, the coefficients 
in eq.\eqref{eq:I2_1} are not the same as in
eq.\eqref{eq:I1_1}, taken in a collinear approximation. It is only for 
keeping the notation as simple as possible that we do not introduce new 
names for the coefficients. In order to diagonalise the
denominator we only have to get rid of $\bk\br$;  the momentum $\bl_c$, 
therefore, does not participate in the shift of momenta (\ref{eq:fushi}),
and the limits of the $\bl_c$ integration  remain the same.
Applying, to \eqref{eq:I2_1}, the shift
(\ref{eq:fushi}) with $\gamma = \mu = 0$, we arrive at  
\begin{align}
\label{eq:coll}
J_2 =& \int d^2 \bk\;d^2\bl \;\Theta(\al \Lambda - |\bl|)\: \frac{A
\,\bk^2 + B \,\bl^2 + C}{[\tau \bk^2
+\omega \bl^2 + R]^4} \nonumber\\ 
=&\; \frac{\pi^2 \bLam^2}{6\,\tau \omega R (R + \bLam^2)}\;\left[ 
A \frac{1}{\tau} + B \frac{1}{\omega}\: \frac{\bLam^2}{R + \bLam^2}
+ C \frac{1}{R}\: \frac{2 R + \bLam^2}{R + \bLam^2} \right]\\[.2cm]
&\mbox{with :} \; \bLam^2 = \omega (\al \Lambda)^2. \nonumber
\end{align}
Below we will show that only the result for $\ep=0$ is needed. In
analogy to eq.\eqref{eq:X} we  define $X_\mathrm{coll}$ to be 
$6 f_\b J_2$  after performing the integration over the transverse momenta:
\begin{align}
\label{eq:X_coll}
\int d^2 \bk\;d^2 \bl \:\ab_\mathrm{coll} \:\Theta(\alpha_\ell \Lambda -
 |\bl_c|) &=  6  \int_0^1 \;\prod_i
d\b_i \;\delta(1-\mbox{$\sum \b_i$})\:\, f_\b
\left.J_2\right|_{\ep=0} \nonumber\\
&= \int \prod_i d\b_i\;\delta(1-\mbox{$\sum \b_i$})\:\; X_\mathrm{coll}.
\end{align}

In the collinear limits all $D_i$, except $D_4$ ($D_3$), become independent 
of $\bl_c$. In the $\b_i$ space these limits therefore read:
\begin{align}
\mathrm{coll}\;{\bar q}:\qquad &\b_{i\ne 4} \quad \ll \quad \b_4 
\label{eq:colllimit4}\\
\mathrm{coll}\;q:\qquad &\b_{i\ne 3} \quad \ll \quad \b_3 
\label{eq:colllimit3}.
\end{align}
However, in  momentum space the integrals of $\ab_\mathrm{coll}$ and $\ab$ have 
different limits of integration.  It turns out that, as a consequence,
$X_\mathrm{coll}$, in the
limit (\ref{eq:colllimit4}) or (\ref{eq:colllimit3}),  can be written  
as a sum of two parts.
The first one coincides with the collinear limit (either (\ref{eq:colllimit4}) or 
(\ref{eq:colllimit3})) of $X$. The other term depends on $\Lambda$,
and vanishes in the collinear limit (\ref{eq:colllimit4}) or 
(\ref{eq:colllimit3}). 

Next we need to consider the soft limit of the collinear limit, the 
collinear-soft approximation $\ab_\mathrm{coll,soft}$. In momentum space, 
this limit is calculated by taking, in
$\ab_\mathrm{coll}(\bk, \bl_c, \al)$, the additional limit 
$\bl_c\sim\al \to 0$ (cf. the definition of $\bl_c$ in (\ref{eq:collvecs})). 
It is then integrated exactly in the same way as just described for
$X_\mathrm{coll}$,
\begin{align}
\label{eq:X_coll,soft}
 \int d^2\bk\, d^2\bl \;\ab_\mathrm{coll,soft} \:\Theta(\alpha_\ell \Lambda -
 |\bl_c|)= \int
 \mbox{$\prod$} \b_i\;\delta(1-\mbox{$\sum \b_i$})\: \; X_\mathrm{coll,soft}. 
\end{align}
Turning to the collinear-soft limit in the $\b_i$ space, one finds that 
the soft limits of the $D_i$ and of the $\left.D_i\right|_\mathrm{coll}$
coincide. However, the $\bl_c$-integration of
$\ab_\mathrm{coll,soft}$ in \eqref{eq:X_coll,soft}, in contrast to 
the integral of $\ab_\mathrm{soft}$, has an upper limit of
integration. 
But one can  show that due to the $\al$-dependence of the 
 integration limit  $\al \Lambda$ in eq.\eqref{eq:X_coll,soft} the position of the
soft limit in the $\{\al, \b_i\}$ subspace (of the $\{\bl_c, \al,
\b_i\}$ space) is  unchanged by the $\bl_c$
integration. Also in case of the integration of $\ab_\mathrm{soft}$
 (without upper limit on $\bl$)  the soft region is unchanged by the
 $\bl$ integration, see  eq.\eqref{eq:softafterb}.
The soft limits of $X$
and $X_\mathrm{coll}$, therefore, are located in the same
region of the $\b_i$ space:
\begin{align}
\label{eq:collsoftlimit}
\mbox{soft of coll:}\qquad \b_2, \b_5, \b_7 \quad \ll \quad \b_1, \b_6, \al
\quad \ll \quad \b_3, \b_4.
\end{align}


$\bullet\quad\int_{s_0} \ab_\mathrm{cen}$

The third type of integral appearing in the real corrections,  
eq.(\ref{eq:sum1CF}) and (\ref{eq:sum1CA}), 
deals with the central approximation
$\ab_\mathrm{cen}$, defined by the limit $\al\to 0$:
\begin{align}
\label{eq:centodo}
J_3 := \int d^{2-2\ep} \bk\;d^{2-2\ep} \bl\:\:\Theta(|\bl|-\al s_0)\:\:
\frac{Z_\mathrm{cen}}{\left[\sum \b_i
\left.D_i\right|_\mathrm{cen}\right]^4}. 
\end{align} 
Its general form reads:
\begin{align}
\label{eq:I3_1}
J_3 = \int d^{2-2\ep} \bk\;&d^{2-2\ep} \bl\:\:\Theta(|\bl|-\al
s_0)\:\: \nonumber\\ 
&\times\frac{\lambda_1\, \bk^2 + \lambda_2\, \bl^2 +  \lambda_3\,
\bk\bl  + \lambda_4\, \bk\br  + \lambda_5\, \bl\br  +
\lambda_6\, \br^2 + \lambda_7}{\left[\tau \,\bk^2 + \rho \,\bl^2 + 2\, \gamma
\,\bk\bl + 2 \eta \,\bk\br + 2 \mu \,\bl\br + D_r \right]^4}.
 \end{align} 
In order to keep the region of the $\bl$-integration as simple as possible, 
we do not perform any shift in $\bl$; this will leave us with angular 
dependent terms of the form $\bl\br$, both in the numerator and in the 
denominator. We only perform the
shift eq.\eqref{eq:fushi} of $\bk$, given by eq.\eqref{eq:fushi},
with $\bl'$ being expressed through $\bl$ and $\br$:
\begin{align}
\label{eq:sub_cen}
\bk &= \bk' - \frac{\gamma}{\tau} (\bl + \lambda \br)
- \frac{1}{\mbox{det} \Omega} \left(\rho \eta - \gamma \mu \right)
\br\:,\quad \lambda = \frac{1}{\mbox{det} \Omega} (\tau\mu - \gamma \eta),
\end{align}
and we obtain
\begin{align}
\label{eq:cen}
J_3 =& \int d^2\bk\; d^2\bl\: \Theta(|\bl|  - \al \sqrt{s_0})
\:\frac{A\, \bk^2 + B\, \bl^2 + C + E \, \bl\br}{[\tau \bk^2 +
\omega (\bl + \lambda \br)^2 + R]^4}\nonumber\\
={}& \frac{\pi^2}{6 \tau \omega R} \;\left( \frac{A}{\tau} \;
  {\mathcal C}_A + \frac{B}{\omega}\; {\mathcal C}_B + \frac{C}{R}\;
  {\mathcal C}_C + \frac{E}{R} \lambda|\br| \; {\mathcal C}_E \right).
\end{align}
The coefficients $\omega$ and $R$ are given by eq.\eqref{eq:omROm}, expressed 
in terms of the coefficients in the denominator of eq.\eqref{eq:I3_1}. Again, we only need the case $\ep=0$. The coefficients in the second line on the rhs 
of (\ref{eq:cen}) are given by
\begin{align}
{\mathcal C}_A  &= \frac{1}{2} \left\{ 1 - \frac{1}{\ka}\Big[\bs -
  (R + \blam^2)\Big] \right\}\nonumber\\
{\mathcal C}_B  &= \frac{1}{2} \left\{\frac{\blam^2 + R}{R} -
  \frac{1}{R \ka} \,\Big[\bs\, (R + \blam^2) - 3 R^2 - \blam^4\Big] \right.\nonumber\\
&\left.\hspace{1cm}-{}  \frac{2}{\ka^3} \, \Big[ (R - \blam^2) (R + \blam^2)^2 + \bs \,(R^2 +
  \blam^4 - 6 R \blam^2)\Big] \right\} \nonumber\\
{\mathcal C}_C &= \frac{1}{2} \left\{ 1 - \frac{1}{\ka} \Big[ \bs + (R
  - \blam^2) \Big] + \frac{2 R}{\ka^3} \Big[ \bs \,(R - \blam^2) + (R +
  \blam^2)^2 \Big] \right\}\nonumber\\
{\mathcal C}_E &= - \frac{1}{2 } \left\{ 1 - \frac{1}{\ka} \Big[ \bs + (R
  - \blam^2) \Big] + \frac{2 R}{\ka^3} \Big[ \bs\, (3 R - \blam^2) + (R +
  \blam^2)^2 \Big] \right\}; \\[.4cm]
&\mbox{where we have defined :\nonumber}
\end{align}
\begin{align*}
&\bs \equiv \omega (\al \sqrt{s_0})^2\:,\qquad \blam^2 \equiv \omega
\lambda^2 \br^2\nonumber\\
&\mbox{and :}\quad \ka \equiv \sqrt{\left(\bs + R + \blam^2 \right)^2 -
  4 \bs \,\blam^2}.
\end{align*}
In complete analogy to the previous integrations we finally define
\begin{align}
\label{eq:X_cen}
 \int d^2\bk\, d^2\bl \;\ab_\mathrm{cen} \:\Theta(|\bl| - \alpha_\ell \sqrt{s_0})= \int
 \mbox{$\prod$} \b_i\;\delta(1-\mbox{$\sum \b_i$})\:  \; X_\mathrm{cen}.
\end{align}

Turning to the soft limit of the central-soft approximation,
we again start in momentum space, introduce Feynman parameters, 
perform the momentum integration, and arrive at $X_\mathrm{cen,soft}$,
in analogy to eq.\eqref{eq:X_cen}. Due to the different limits of integration 
of $\ab$ and $\ab_\mathrm{cen}$, only a part of
$X_\mathrm{cen}$ matches the central approximation of $X$, the rest
depends on $s_0$ and vanishes in the central limit. Similar to
the collinear approximation,  the $\al$-dependence of the
lower limit on the $\bl$ integration \eqref{eq:X_cen} has the effect that the momentum
integration does not change the position of the soft region in the
$\{\al, \b_i\}$ space. The soft limit is located in the region:
\begin{align}
\label{eq:censoftlimit}
\mbox{soft of cen:}\qquad \b_2, \b_5, \b_7 \quad \ll \quad \al \quad
\ll \quad \b_1, \b_3, \b_4, \b_6.
\end{align}


As we have mentioned before, when treating individual diagrams, $\ab$, 
additional logarithmic divergences appear which do not show up in the sum.  
They occur in the  following limits:
\begin{align}
\luv \;&:\; |\bl|\to \infty\nonumber\\
\mbox{is}\;& :\; |\bl|\sim \sqrt \al \to 0\nonumber \\
\mbox{lr}\;& :\; \bl \to \br.  
\end{align}
Tab.\ref{tab:ABdiv} contains, in the third columns, a list of those diagrams 
where these divergences appear. These divergences lead to additional 
subtractions which we have to describe in some detail. 
Let us start with the $C_F$ parts. They do not contribute
to the central limit. The only additional
divergence is of the type  $\luv$; 
in this  limit, all denominators $D_i$
(\ref{eq:D1}) - (\ref{eq:D7}) are proportional to $\bl^2$, 
except for $D_2$ which is independent of $\bl$. Any  $\ab$ containing  
$D_2^2$ in the denominator and a term proportional to $\bl^2$ in the
numerator will therefore, in the  limit $\bl
\to \infty$, be proportional to $1/\bl^2$, leading to an UV divergent $\bl$
integration. Subtracting from such diagram its approximation in this
UV-limit, $\ab_{\luv}$, would cancel the $\luv$ divergence. However, the $\bl$
integration then becomes IR divergent instead, since
$\ab_{\luv}\sim 1/\bl^2$. 
We therefore define our subtraction
term in the Feynman parameter space where it is easy to avoid this
additional IR divergence.
Let us demonstrate this at a simplified expression
$A=1/[ (c+\bl^2)D_2 ]$.  $c$ is a constant and $D_2$ does not depend on
$\bl$. We introduce the Feynman representation and integrate over
$\bl$ (assuming already a term that cancels the $\luv$ divergence):
\begin{align}
\label{eq:luvex}
&\int_0^\infty d\bl^2\: A = \int_0^\infty d\bl^2\frac{1}{(c + \bl^2) D_2} =
\int d\b_1 d\b_2  \int_0^\infty d\bl^2
\frac{\delta(1-\b_1-\b_2)}{[\b_1(\bl^2 + c) + \b_2 D_2]^2} \nonumber \\
&=\int d\b_1 d\b_2 \;\delta(1-\b_1-\b_2) \:\frac{1}{\b_1(\b_1 c + \b_2
  D_2)}=\int_0^1 \frac{d\b_1}{\b_1(\b_1 (c - D_2) + D_2)}
\end{align}
The divergence at $\bl\to \infty$  appears in the limit $\b_1\to 0$
after the integration. The natural subtraction in the Feynman
parameter space therefore reads
\begin{align}
\label{eq:naturalsub}
\int_0^1 \left(\frac{d\b_1}{\b_1(\b_1 (c - D_2) + D_2)} -
\frac{d\b_1}{\b_1 D_2}\right).
\end{align}
As an alternative
subtraction term we could use the
$\luv$ limit of $A$, which is $1/[\bl^2 D_2]$, in Feynman parameter
representation. It is obtained  by
taking the integrand in the second line of eq.\eqref{eq:luvex} in the
limit $\b_1\to 0$ excluding the argument of the $\delta$-function
from this approximation. Carrying out the $\b_2$ integration this results in 
\begin{align*}
\frac{1}{\b_1(1-\b_1) D_2}.
\end{align*}
The additional divergence of this term at $\b_1\to 1$ corresponds to
the limit $\bl \to 0$ in the momentum space and can be avoided by
using the subtraction eq.\eqref{eq:naturalsub}.

We therefore regularise the $\luv$ divergence of our diagrams in the
following way.
Due to the behaviour of the $D_i$'s in the $\luv$ limit mentioned
above, the region $\bl\to \infty$ translates to the region
$\b_{i\ne2}\ll \b_2$ in the Feynman parameter space. Each $\luv$
divergent diagram has a $D_2$ in the denominator. After the introduction
of the Feynman representation and after carrying out
the momentum integration we always
perform  the $\b_2$ integration by means of the $\delta$-function in
case of these diagrams. The subtraction term is then determined by
taking the limit $\b_{i\ne 2}\ll 1$ in $2-2\ep$ transverse dimensions:
\begin{align}
\label{eq:luvreg}
\int d^{2-2\ep} \bk\;d^{2-2\ep} \bl \:\ab
&= \int  d\b_2\, \prod_id\b_i\;\delta(1-\b_2-\mbox{$\sum \b_i$})\:\; X
= \int \prod_i d\b_i\; \tilde{X}\nonumber\\
&= \int \prod_i d\b_i\;\left( \tilde{X} -
\tilde{X}_\mathrm{\luv} \right) + \int \prod_i d\b_i\;
\tilde{X}_\mathrm{\luv}, \\
&{}\qquad \mbox{with:}\quad\tilde{X}_\mathrm{\luv} = \left.\tilde{X}\right|_{\b_i\ll 1}\nonumber
\end{align}
To perform the momentum integration we needed to calculate 
the  integral in eq.\eqref{eq:full} for $\ep\ne 0$.
The tilde in eq.\eqref{eq:luvreg} indicates that $\b_2$ in
$X$ is expressed in terms of the
other $\b_i$. 
We perform the $\b_i$ integration in the re-added piece analytically,
obtaining  an $\ep$-pole and  finite terms. 
Summing up the re-added $\luv$ subtractions from all diagrams we find
that the poles cancel, as expected.

We can now write down the final result (after momentum integration) 
for the $C_F$ part of the real corrections
\eqref{eq:sum1CF}. Using the definitions
(\ref{eq:X}), (\ref{eq:X_coll}), and (\ref{eq:X_coll,soft}) we obtain
\begin{align}
\label{eq:sum2CF}
 \left.  \Phi_{\gs}^{(1,{\mathrm real})}\right|_{C_F}^{\mathrm
 finite}&= \frac{e^2e_f^2 \sqrt{N_c^2-1} }{ (2 \pi)^6} \Bigg\{ 
\sum_{C_F}\; \int_0^1 d\alpha\int_0^{1-\alpha} 
 \frac{d\alpha_\ell}{\al}
 \int_0^1  \mbox{$\prod$} d\b_i\:\delta(1-\mbox{$\sum$}\b_i)
 \nonumber\\
&\hspace{-.7cm} 
 \bigg( X - X_\mathrm{soft} - X_{\luv} - (X_q - X_{q,\mathrm{soft}}) -  (X_{\bar q} -
 X_{{\bar q},\mathrm{soft}}) \bigg)^{C_F} \nonumber \\
&+\frac{\pi^2}{24}\;(3 \pi^2 - 28) \Bigg\}.
\end{align}
The superscript indicates that only the $C_F$ part of the bracket is 
taken, excluding the color factor $C_F$.
To simplify notations, we have written the delta function 
$\delta(1-\mbox{$\sum$}\b_i)$ also for the $X_{\luv}$ term. 
However, as discussed above, $\b_2$ in $X_{\luv}$ is understood 
to be already expressed in terms of the other $\b_i$.  We have  set
$\ep=0$ in \eqref{eq:sum2CF} since the $\b_i$ integrals are convergent
by construction.  This is why we  needed to calculate the integrals over
$\ab_\mathrm{coll}$ and $\ab_\mathrm{cen}$ only for the case $\ep=0$.
The term in the third line of eq.\eqref{eq:sum2CF} results from the sum of
the re-added $\luv$ subtractions.

Now we turn to the $C_A$ parts of the diagrams. We have to deal with
collinear divergences. However, according to \cite{real} the collinear
approximations of all $C_A$ parts sum up to zero; we can therefore
just subtract from the divergent $\ab's$ their collinear approximation
without re-adding it (as in the $C_A$ case, we restrict the
integration to a cone since otherwise it would be UV divergent).

According to tab.\ref{tab:ABdiv}, in the $C_A$ parts we encounter all three 
types of additional divergences.
As to the is(intersoft)-divergence, one finds in momentum space:  
\begin{align*}
\sum_{C_A}\,(\ab - \ab_\mathrm{soft})_{is} = 0.
\end{align*}
We can therefore regularise this divergence by subtracting, from 
$(\ab-\ab_\mathrm{soft})$, its is-approximation, without re-adding it. 
In the appendix we provide the combinations 
$(\ab - \ab_\mathrm{soft})_{is}$ in the $\b_i$
space, $(X-X_\mathrm{soft})_\mathrm{is}$, which can be obtained from  
$X-X_\mathrm{soft}$ by taking the
is-limit (before making use of the $\delta$-function): 
\begin{align}
\label{eq:islimit}
\mathrm{is :}\qquad\b_2, \b_5, \b_7, \al \quad \ll \quad \b_1, \b_3, \b_4, \b_6.
\end{align}

In the  lr-limit ($\bl\to\br$), it is the central approximation of certain
diagrams that diverges.  Let us recall the  structure of the $\A\A$ and $\A\B$ 
diagrams in the central limit $\al\to 0$ (see \cite{real}):
\begin{align}
\label{eq:cenAA}
\left. \A\A_{ij}\right|_{\mathrm{cen}} &= \frac{\alpha_1^3
\,\alpha_2^3 \,C_A}{(\bl - \br)^2}\;\frac{1}{D^2(\bk+\br)}\;D_{ij}\;,\\
\label{eq:cenAB}
\left. \A\B_{ij}\right|_{\mathrm{cen}} &= \frac{- \alpha_1^3\, \alpha_2^3
  \,C_A}{(\bl - \br)^2}\; \frac{1}{D(\bk+\br) D(\bk+\bl)}  \;D_{ij}\;.
\end{align}
where $D(\bk) = \alpha_1 \alpha_2 + \bk^2$.  The matrix $D_{ij}$ reads :
\begin{equation}
\nonumber
  \begin{pmatrix}
    0 & 0 & 0 & -\frac{1}{2} & \frac{\bm r^2}{\bm\ell^2}\\
    0 & 0 & 0 & -\frac{1}{2} & \frac{\bm r^2}{\bm\ell^2}\\
    0 & 0 & 0 & 0 & 0 \\
    -\frac{1}{2} & -\frac{1}{2} & 0 
      & 2 & 0\\ 
    \frac{\bm r^2}{\bm\ell^2}& \frac{\bm r^2}{\bm\ell^2} & 0 & 0 & 0
\end{pmatrix}\;.
\end{equation} 
The denominators in eqs.(\ref{eq:cenAA}) and (\ref{eq:cenAB}) correspond
to :
\begin{align}
\label{eq:Di_cen}
\left.D_2\right|_\mathrm{cen}&=D(\bk+\br)\nonumber\\
\left.D_5\right|_\mathrm{cen}&=\a_1 \a_2 (\bl-\br)^2\\
\left.D_7\right|_\mathrm{cen}&=D(\bk+\bl)\,.\nonumber
\end{align}
Eqs.(\ref{eq:cenAA}) and (\ref{eq:cenAB}) show that, if a diagram
contributes to the central limit, its approximation in this limit is
divergent as $\bl\to \br$. However, for a given pair 
(i,j),  eq.\eqref{eq:cenAA} and  eq.\eqref{eq:cenAB} become equal with
opposite sign as $\bl \to \br$, since
$\left.D_7\right|_\mathrm{cen}\to\left.D_2\right|_\mathrm{cen}$. To
regularise the $\luv$ divergence we
therefore choose a common set of $\{\beta_i\}$ for $\A\A_{ij}$ and $\A\B_{ij}$ in the
following way:
\begin{align}
\label{eq:lr_param}
\A\A_{ij}\sim\frac{1}{D_xD_5D_2^2} &= \int
\{d\b_i\}\;\frac{\delta(1-\mbox{$\sum$}\b_i)}{[\b_xD_x+\b_5D_5+(\b_2+\b_7)D_2]^4},\\
\A\B_{ij}\sim\frac{1}{D_xD_5D_2D_7} &= \int
\{d\b_i\}\;\frac{\delta(1-\mbox{$\sum$}\b_i)}{[\b_xD_x+\b_5D_5+\b_2D_2+\b_7D_7]^4}\,.\nonumber
\end{align}
The momentum integration is performed in the same way as for all other 
diagrams. For the
$\beta_i$ integration, we consider the sum
($\A\A_{ij}+\A\B_{ij}$).  Our parameterisation \eqref{eq:lr_param}
ensures that the  lr-divergence (in the $\b_i$ space)
cancels between the two diagrams.  All what we have said about the 
lr-divergence in the $\A\A_{ij}$ and $\A\B_{ij}$ in an analogous
way also applies to the $\B\B$ and $\B\A$ diagrams. 
The parameterisation of $\B\B_{ij}$ is given by
eq.\eqref{eq:lr_param} with $D_2$ being replaced by $D_7$. Note that it is 
only those diagrams that are of type $\A\A$ or $\B\B$ and contribute to the
central limit which need this special parameterisation.

The matrix $D_{ij}$ reveals another potential divergence: $\left.\A\A_{15}\right|_\mathrm{cen}$
and $\left.\A\A_{25}\right|_\mathrm{cen}$ are obviously divergent in the soft limit. Their
soft approximations which we will subtract are proportional to $1/\bl^2$
giving rise to an UV divergent $\bl$ integration. However, one can
show that this divergence is automatically cancelled by
$\left.\A\A_{15}\right|_\mathrm{is}$ and
$\left.\A\A_{25}\right|_\mathrm{soft}$, respectively.

Following \eqref{eq:sum1CA}, we have to subtract the central and
central-soft approximations in the region where  $|\bl|>\al
\sqrt{s_0}$.  Eqs.(\ref{eq:cenAA}) and (\ref{eq:cenAB})
show  that the central approximations of the diagrams 14,24,41,42,44
are also $\luv$ divergent. In order to regularise this $\luv$ divergence in
the same way as discussed above one would need to calculate an  
$\bl$-integral  with non-zero lower limit in arbitrary
dimensions. This can be avoided since, according to the structure of $D_{ij}$, 
$\sum_{14,24,41,42,44}\,\ab_\mathrm{cen} =0$; 
we are therefore free to subtract the central approximation for
these diagrams in the whole momentum space ($|\bl|>0$) translating
to:  $X-X_{\al=0}$. The
$\luv$ divergence then is 
regularised by subtracting the $\luv$ approximation of this
combination, $(X- X_{\al =0})_{\luv}$, after doing the $\b_2$
integration by means of the $\delta$-function. Since the central and $\luv$
approximations commute, this is equal to  $(X_{\luv}-
\left.X_{\luv}\right|_{\al =0})$. We re-add these
subtractions, carry out the remaining $\b_i$-integrations and find the
expected cancellation of the $\ep$ poles. 

Our final result for the $C_A$ part, defined in eq.\eqref{eq:sum1CA}, 
then reads:
\begin{align}
\label{eq:sum2CA}
& \left.  \Phi_{\gs}^{(1,{\mathrm real})}\right|_{C_A}^{\mathrm
 finite}= \frac{e^2e_f^2 \sqrt{N_c^2-1}  }{ (2 \pi)^6} 
  \Bigg\{ \nonumber\\
&\hspace{.6cm}{} \sum_{12,14,24,44}^* \int_0^1 d\alpha \int_0^{1-\alpha}
 \frac{d\alpha_\ell}{\al} \int_0^1 \mbox{$\prod$}d\b_i\:\delta(1-\mbox{$\sum$}\b_i) \nonumber \\
&\bigg(  
 X  -  X_{\luv} - (X  -  X_{\luv})_{\al=0} - X_\mathrm{soft} - \left( X_\mathrm{coll} - X_\mathrm{coll,soft} \right) 
  \bigg)^{C_A}\nonumber \\
&\hspace{.6cm}{}+ \sum_{rest}^* \int_0^1 d\alpha \int_0^{1-\alpha}
 \frac{d\alpha_\ell}{\al} \int_0^1 \mbox{$\prod$}d\b_i \:\delta(1-\mbox{$\sum$}\b_i)\nonumber \\
&\bigg(X  -  X_\mathrm{soft}  -
\left( X_{\mathrm{cen}} - X_\mathrm{cen,soft} \right) - \left( X -
 X_\mathrm{soft} \right)_{is}- \left( X_\mathrm{coll}-
 X_\mathrm{coll,soft} \right)
   \bigg)^{C_A} \nonumber\\
&\hspace{.6cm}{}- \sum_{C_A}^* \int_0^1 d\alpha \int_{1-\alpha}^{\infty}
 \frac{d\al}{\al} \int_0^1
 \mbox{$\prod$}d\b_i\:\delta(1-\mbox{$\sum$}\b_i) \: X^{C_A}_\mathrm{cen}\nonumber\\
&\hspace{.6cm}{}-\frac{\pi^2}{48}\;(3 \pi^2 - 28)  \Bigg\}\,.
\end{align}
 The summation in the second line is understood to extend over $\A\A,
\A\B, \ldots$ and to include also 21,41,42; by 'rest' we mean all other
$C_A$ diagrams.
The * on the sum  indicates that in case of an lr-divergent diagram we
are not allowed to separate the sum of $\A\A$ and $\A\B$ 
($\B\B$ and $\B\A$) diagrams . The term in the last line 
again is a result of the re-added $\luv$ subtractions.

Let us summarise the procedure of deriving, from the $\ab's$ given
in \cite{real}, the Feynman para\-meter
representation of the real corrections written in eqs.(\ref{eq:sum2CF}) and 
(\ref{eq:sum2CA}). The results of \cite{combine}, repeated in 
our eqs.(\ref{eq:sum1CF}) and (\ref{eq:sum1CA}), define, for the sum of all
diagrams, the necessary subtractions of collinear and soft divergences    
and of the central region. We then turn to individual diagrams $\ab$:
we start from momentum space and calculate the various approximations required 
in (\ref{eq:sum1CF}) and (\ref{eq:sum1CA}). We then introduce the 
Feynman parameter representation, written in (\ref{eq:afterb})
and  (\ref{eq:xafterb}), retaining still the transverse momenta to be 
integrated over. In case of the collinear
limit we change the integration variable $\bl$ to $\bl_c$. 
In the integrands, we explicitly write the squares and scalar products 
of transverse momenta. For each approximation we then obtain two sets of
coefficient functions, $\{\lambda_i\}$ in the numerators and   
$\{\tau, \rho, \gamma, \ldots, D_r\}$ in the denominators of the integrand,
respectively. From the latter set we calculate the quantities $\omega,
R, \lambda$ in (\ref{eq:omROm}) and (\ref{eq:sub_cen}), and we define 
the appropriate shifts of the transverse momenta. Applying
these shifts to the numerator gives the coefficients $A, B, C,
E$. Now, we do the integration over transverse momenta, observing 
the necessary limits of integration. The results,
eqs.(\ref{eq:full}), (\ref{eq:coll}), and  (\ref{eq:cen})
are expressed in terms of the coefficient functions obtained before.   
Using the definitions
(\ref{eq:X}), (\ref{eq:X_coll}), and (\ref{eq:X_cen}) we are finally left with
a set of expressions $X, X_\mathrm{cen}, X_\mathrm{coll},
X_\mathrm{soft}, \ldots $
which are functions of the momentum fractions $\a, \al$, of the Feynman
parameters $\{\b_i\}$, of the cutoff parameters $\Lambda$ and 
$s_0$, and of the virtuality of the t-channel gluons, $\br^2$.
After calculating $X_{\luv}$ and $\left(X-X_\mathrm{soft} \right)_{is}$ 
we have got everything that in eqs.(\ref{eq:sum2CF}, \ref{eq:sum2CA}) 
is needed for a finite integral representation of the single diagram, 
$\ab$. 

We have implemented this procedure into a MATHEMATICA
program, and we have applied it to  all $\ab's$. The result are analytic
expressions for all the  $X, X_\mathrm{cen}, X_\mathrm{coll}, \ldots $
needed in eqs.(\ref{eq:sum2CF}) and (\ref{eq:sum2CA}). They are contained 
in a set of files described in the appendix.


\section {Numerical results}
For the final integrals in eqs.(\ref{eq:sum2CF},\ref{eq:sum2CA}) we
have carried out the remaining integrations 
over $\a, \al, \b_i$ numerically. 
We have done this for each $\ab$ separately, except for 
the lr-divergent diagrams, where we have considered the combination
$\A\A_{ij}+\A\B_{ij}$ and $\B\B_{ij}+\B\A_{ij}$, respectively. We have 
used the Monte-Carlo routine VEGAS. 
For each diagram the integrand is given in a FORTRAN code which is 
described in the appendix. As a result, we have obtained values of
\begin{align*}
\left.\Phi_{\gs}^{(1,{\mathrm real})}\right|_{C_A, C_F}^{\mathrm
finite}\:,\;\quad\mbox{as a  function of }\;\br^2, \Lambda \:\mbox{and}\: s_0. 
\end{align*}

To be precise, we start from the expression \eqref{eq:phi1} of the NLO 
corrections to the $\gs$ impact factor, and we combine the finite real 
corrections in the fourth line with the $\Lambda$ and $s_0$-dependent pieces 
in the second and third line:  
\begin{align}
\label{eq:phi1p}
  \Phi_{\gs}^{(1)} = &
  \left.\Phi_{\gs}^{(1,{\mathrm virtual})}\right|^{\mathrm finite}
  - \frac{2 \Phi_{\gs}^{(0)}}{(4\pi)^2} \left\{
    \beta_0 \ln\frac{\br^2}{\mu^2} + C_F\ln(\br^2) 
  \right\}
\nonumber\\
&+\int_0^1 d\alpha\,\int\,\frac{d^2\bk}{(4\pi)^2}\,{\mathcal{I}}_2(\alpha,\bk)\,\bigg\{
 - C_A \ln^2 M^2\nonumber\\
  &\qquad\quad + 2C_F \left[8 - 3\ln \alpha(1-\alpha) + \ln^2 M^2
  + \ln^2 \frac{\alpha}{1-\alpha} \right]
\bigg\} + \Phi_{\gs}^{\mathrm real}
\end{align}
with
\begin{align}
\Phi_{\gs}^{\mathrm real} = &\;
  C_A \left.\Phi_{\gs}^{(1,{\mathrm real})}\right|_{C_A}^{\mathrm finite}
  + C_F \left.\Phi_{\gs}^{(1,{\mathrm real})}\right|_{C_F}^{\mathrm finite} - 3\,C_F\,\frac{\Phi_{\gs}^{(0)}}{(2\pi)^2}\; \ln
  \Lambda\nonumber\\
+&\,C_A\,\int_0^1
  d\alpha\,\int\,\frac{d^2\bk}{(4\pi)^2}\,{\mathcal{I}}_2(\alpha,\bk)\,\ln^2\alpha(1-\alpha)  s_0 
\label{eq:realp1}\\[.2cm]
=&\,e^2 e_f^2 \;( \Phi_A + \Phi_F + \Delta_\Lambda + \Delta_{s_0})\,.
\label{eq:realp2}
\end{align}
In eq.\eqref{eq:realp2} we have introduced short hand 
notations for the four terms in eq.\eqref{eq:realp1}. 
The LO impact factor is given by $\Phi_{\gs}^{(0)}
= \int d^2\bk \int_0^1 d\alpha \, {\mathcal{I}}_2(\alpha, \bk)$, with
$\mathcal{I}_2$ being defined in eq.\eqref{eq:bornifl}.  The
integration  of $\Delta_\Lambda, \Delta_{s_0}$ is straightforward and we obtain
$\Phi_{\gs}^{\mathrm real}$ as a function of $\br^2, \Lambda$ and
$s_0$. 

In our numerical analysis we restrict ourselves to two questions.  
First we investigate the $\Lambda$ dependence of the real
corrections. The cutoff parameter $\Lambda$ specifies the region of the
subtraction of the collinear approximations. Since these subtracted
terms are re-added, the NLO corrections to the impact factor must be
independent of $\Lambda$. This must also be true for 
$\Phi_\gs^{\mathrm real}$, since it
contains all $\Lambda$ dependent terms. 
$\Delta_{s_0}$ in eq.\eqref{eq:realp2} does not depend on $\Lambda$,
so $\Phi_F + \Delta_\Lambda$ and $\Phi_A$ must be $\Lambda$-independent 
separately. 

Second, we study the $s_0$ dependence of the real corrections. 
Since the entire $s_0$ dependence of the NLO corrections is contained in the 
real corrections and in the finite term (cf.eqs.(\ref{eq:realp1}, 
\ref{eq:realp2}), the results obtained in the present paper already 
allow to study the scale dependence of the NLO $\gs$ impact factor.
Whereas the complete scattering amplitude, involving the NLO impact factors   
and  the NLO BFKL Green's function, has to be invariant under changes   
of $s_0$, the impact factor alone is expected to vary. Since a decrease 
of $s_0$ in the energy dependence $(\frac{s}{s_0})^{\omega}$ will enhance 
the scattering amplitude, the combined $s_0$ dependence of the impact factors 
and of the BFKL Green's function has to compensate this growth. We find that
the $s_0$-dependent part of the NLO corrections to the $\gamma^*$ impact 
factor has the sign opposite to the LO impact factor and, in absolute value, 
becomes more significant when $s_0$ becomes smaller. As a result, 
this part of the NLO corrections, in fact, tends to make the LO impact factor 
smaller.   
    
\begin{figure}[h]
\begin{center}
 \includegraphics[width=10cm]{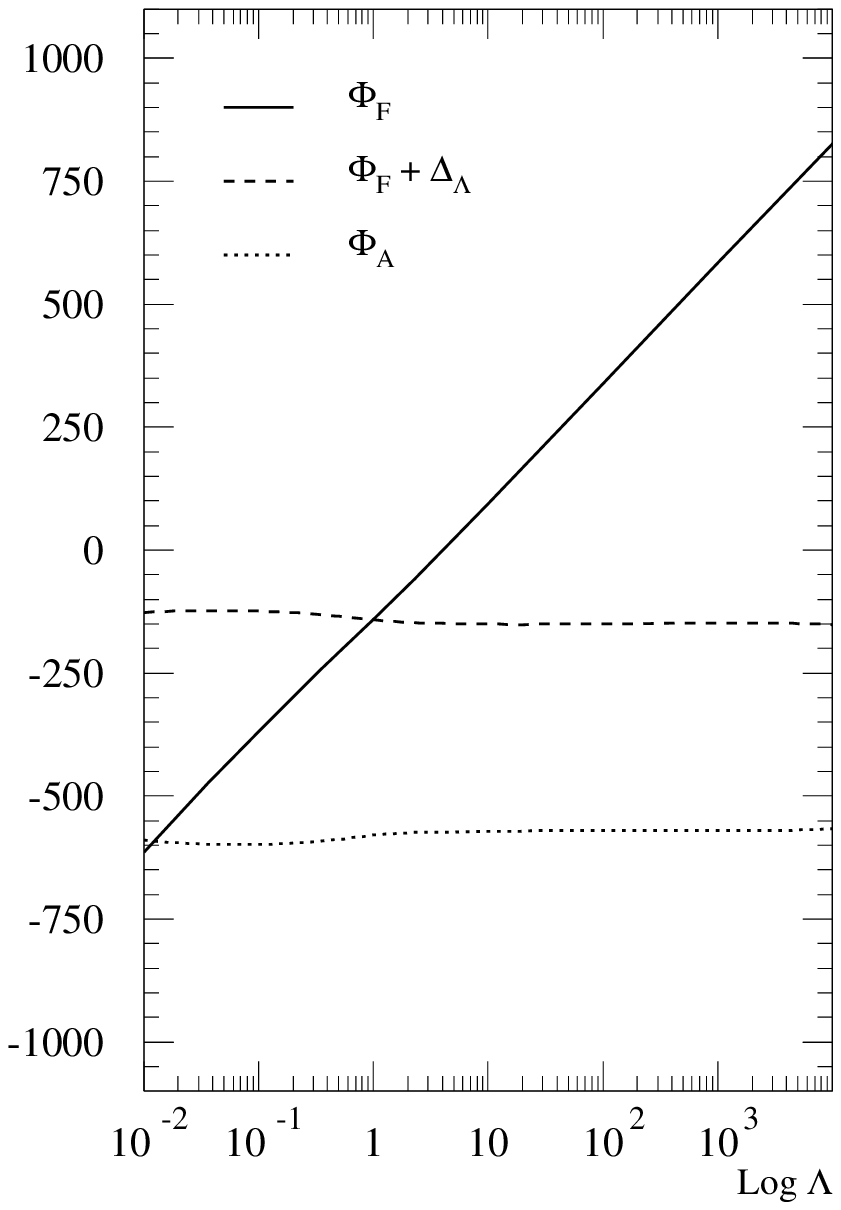}
\caption{\label{lambda-dep} The dependence of the real corrections on
  $\Lambda$ (with  $\br^2 = s_0 = 1$) }
\end{center}
\end{figure}

Fig.\ref{lambda-dep} shows the
dependence of the $\Lambda$ dependent parts of $\Phi_\gs^{\mathrm real}$  on
 the momentum cutoff $\Lambda$ for $\br^2 = s_0 = 1$ and $N_c=3$. 
According to eq.\eqref{eq:realp1} the only $C_A$ term in $\Phi_{\gs}^{\mathrm
real}$ is $\Phi_A$. It includes $\Lambda$ dependent terms, since, as
discussed above, we also performed collinear subtractions for the $C_A$ parts 
of the diagrams. Fig.\ref{lambda-dep} shows that, in fact, the $\Lambda$
dependence is very weak.  As to the $C_F$ terms, $\Phi_F$ turns out to be 
proportional to $\ln \Lambda$. This growth with $\Lambda$, however, 
is fully compensated by $\Delta_\Lambda$. 
Note that $\Phi_F$ is a sum of many Feynman diagrams, 
whereas $\Delta_\Lambda \sim \ln \Lambda$. The compensation of
the $\Lambda$ dependence, therefore, represents a rather stringent test
of the calculation of the $\gs$ impact factor.

\begin{figure}[h]
\begin{center}
${}$\vspace{-.5cm}
 \includegraphics[width=12cm]{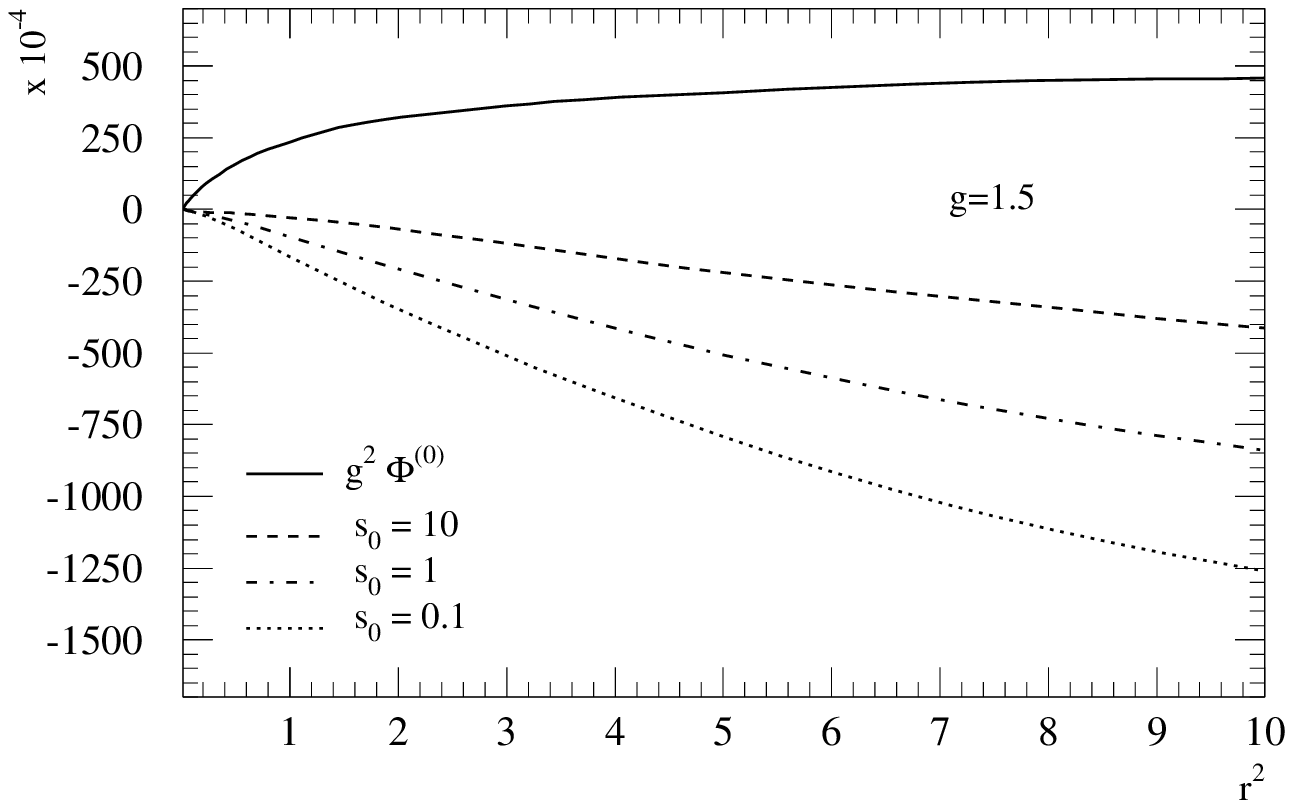}
\caption{\label{ifac_1} $\Phi'_\gs $ at different different values of
$s_0$ }
${}$\vspace{-.5cm}
  \includegraphics[width=12cm]{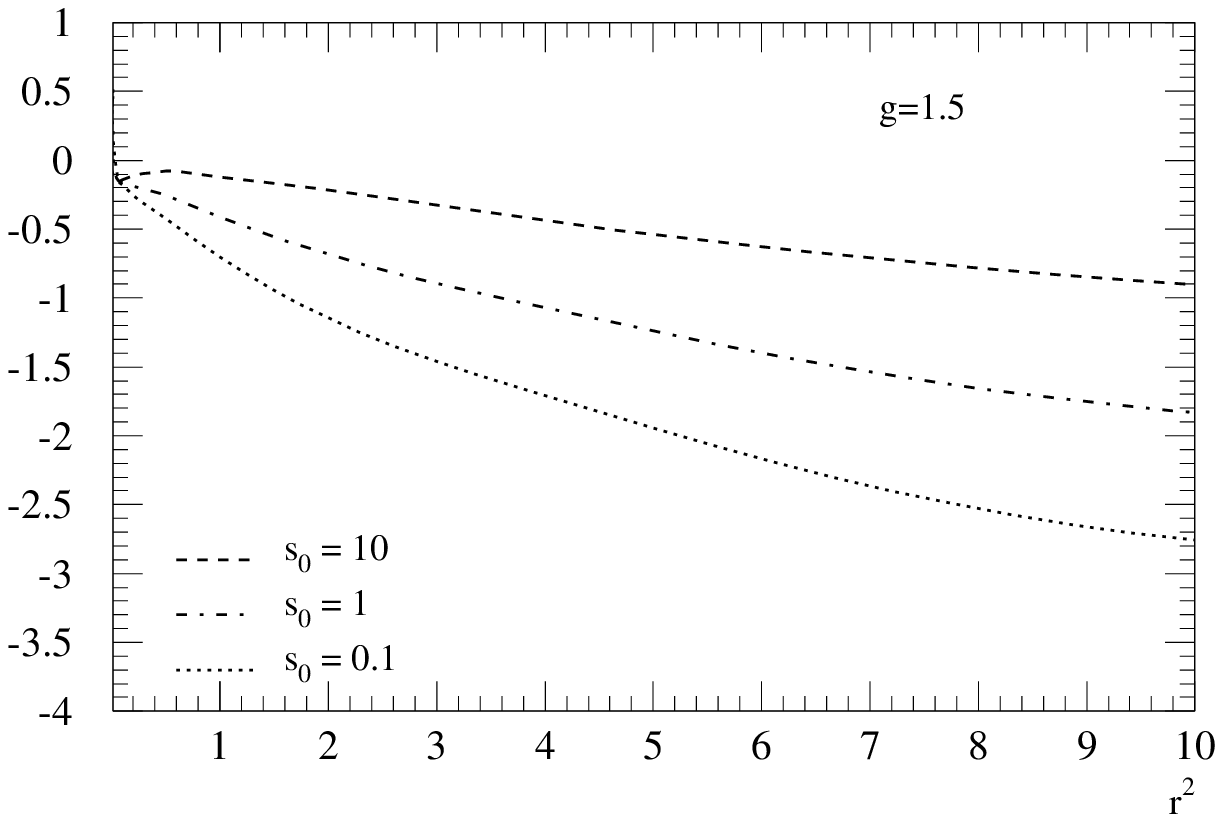}
\caption{\label{ifac_2} The ratio $\Phi'_\gs /(g^2
  \Phi^{(0)}_\gs) $ at different $s_0$ }
\end{center}
\end{figure}

Next we address the dependence of the NLO $\gs$ impact factor on the
energy scale $s_0$.
The full LO and NLO impact factor can be written as 
$\Phi_\gs = g^2\Phi_\gs^{(0)} +
g^4\Phi_\gs^{(1)}$, where $g^2=4\pi \as$. Since, at the moment, we only know 
the real corrections, we compute, as a part of the full answer:  
\begin{align*}
\Phi'_\gs = g^2\Phi_\gs^{(0)} + g^4\Phi_\gs^{\mathrm
real}.
\end{align*}
We  set $e^2e_f^2=1$. For the photon virtuality we choose,
as a typical value in $\gs\gs$ scattering, $Q^2=15$ GeV$^2$. 
This choice affects the strong coupling, $\as(Q^2)=0.18$ or $g=1.5$,
and, through (\ref{norm}), the normalization of $\br^2$ and $s_0$. 
Fig.\ref{ifac_1} compares $\Phi'_\gs$ to the LO impact factor
$g^2\Phi_\gs^{(0)}$ as a function of $\br^2$ at different values of
$s_0$. 
In agreement with gauge invariance, both the LO impact factor and the
real corrections vanish at $\br^2=0$.  
The ratio of $\Phi'_\gs$ and  $g^2\Phi_\gs^{(0)}$ is shown in
fig.\ref{ifac_2}. The real corrections are negative and rather
large. The overall magnitude is not of much significance, 
because we have considered only a part of the NLO corrections. More
important
is the fact that, in absolute terms, $\Phi'_\gs$ becomes more significant 
for smaller values of $s_0$. Since we
included all $s_0$ dependent terms in $\Phi'_\gs$, this implies that
the $\gs$ impact factor tends to become smaller with decreasing $s_0$. This
behaviour goes in the expected direction.


\section{Conclusion}
In two previous publications \cite{real,combine}, 
the real corrections to the $\gs$ impact
factor have been computed, and in \cite{combine} infrared finite 
combinations of real and virtual corrections have been obtained. 
The resulting integrals still contain the integration over the 
transverse momenta. In the present paper   
we have carried out this integration for the real corrections,
restricting ourselves to the case of longitudinal photon
polarisation. To allow for further theoretical analysis we have 
introduced Feynman parameter representations, and we have performed the 
integration over the transverse momenta analytically. Our results  
are finite Feynman parameter integral expressions for each
diagram. These expressions which may serve as a basis for future 
studies are presented in a computer code, which we describe in an appendix.   

In a first exploratory study we then have evaluated these integrals
numerically. Two questions have been addressed: first, we have shown that
the NLO corrections to the impact factor are independent of the
parameter $\Lambda$ specifying the region of the collinear
subtraction. Second, we have studied the dependence upon the 
energy scale $s_0$. A physical scattering amplitude (e.g. for the 
$\gamma^*\gamma^*$ scattering process), when consistently evaluated in NLO 
accu\-racy, has to be invariant under changes of $s_0$. 
The NLO impact factor, which is part of the scattering amplitude,
will change when $s_0$ is modified. In our analysis,   
the NLO impact factor was found to be negative and to become 
more significant when the value of the 
energy scale $s_0$ is lowered. This is, at least, consistent with the 
general expectation.  

What remains to be done for the real corrections, 
is the extension to the case of the transverse photon polarisation. 
It still requires some efforts, 
as there are additional divergences to be dealt with. 
However, with the tools developed in the longitudinal case and described in 
this paper, we hope to fulfill this task in the near future.  


\appendix*
\section{}

In this appendix we describe  our MATHEMATICA and
FORTRAN files that  contain the analytic expressions of our diagrams
 and are available at:
\begin{center}
http://www.desy.de/uni-th/smallx/kyrie/ImpFactor/index.html
\end{center}

{\bf A.\hspace{.5cm}}     
First, we provide the products of diagrams in momentum space. The
$\ab$'s, defined in eq.\eqref{eq:def_ab}, are listed  in  six
2-dim. MATHEMATICA arrays, corresponding
to the three groups  of products, $\A\A_{ij}, \A\B_{ij}, \B\B_{ij}$
and the two color factors $C_A, C_F$. With $\A\A_{ij}=C_A
\;\A\A_{ij}^{C_A} + C_F \;\A\A_{ij}^{C_F}$ (and correspondingly
$\A\B_{ij}, \B\B_{ij}, \ldots$) we define the following arrays:
\begin{align*}
&\mathtt{M\_CFaa[[ i, j ]]}= \frac{(Q^2)^3}{\ae \ake} \A\A_{ij}^{C_F}\,,\quad
&\mathtt{M\_CAab[[ i, j ]]}= \frac{(Q^2)^3}{\ae \ake} \A\B_{ij}^{C_A},
\end{align*}
and  accordingly
\begin{align*}
&\mathtt{M\_CFab[[ i, j ]]}\,,\quad \mathtt{M\_CFbb[[ i, j ]]}\,,\quad
&\mathtt{M\_CAaa[[ i, j ]]}\,,\quad \mathtt{M\_CAbb[[ i, j ]]}.
\end{align*}
The six arrays are stored in the files
\begin{center}
 $\mathtt{M\_CFaa.m},\:\: \mathtt{M\_CFab.m},\:\:
\mathtt{M\_CFbb.m}\quad$ and $\quad\mathtt{M\_CAaa.m}, \:\:\mathtt{M\_CAab.m},
 \:\:\mathtt{M\_CAbb.m}$
\end{center}
In these arrays the following symbols are used:
\begin{align*}
\mathtt{L2}=\bl^2,\quad\mathtt{K2}=\bk^2,\quad\mathtt{R2}=\br^2,\quad&\mathtt{KL}=\bk\bl,\quad\mathtt{KR}=\bk\br,\quad\:\mathtt{LR}=\bl\br,\\
\mathtt{ep}=\ep,\quad\mathtt{a}&=\a,\quad\mathtt{al}=\al,
\end{align*}
$\mathtt{D1}\ldots\mathtt{D7}$ are the denominators as given in
eqs.(\ref{eq:D1}-\ref{eq:D7}). The $\B\A$ products are obtained via
$\mathtt{M\_CFba[[ i, j ]]}=\mathtt{M\_CFab[[ j, i ]]}$ and 
$\mathtt{M\_CAba[[ i, j ]]}=\mathtt{M\_CAab[[ j, i ]]}$.

{\bf B.\hspace{.5cm}} We also provide the analytic expressions  $X,
X_\mathrm{soft}, \ldots$, used in the formulae for
the real corrections in eqs.(\ref{eq:sum2CF}) and (\ref{eq:sum2CA}). For
any diagram a set of expressions $X,
X_\mathrm{soft}, \ldots$ is needed.  In accordance with our labelling of the
diagrams, we use subscripts for the X's and for their approximations. 
For instance, $[X_\mathrm{soft}]^{\mathrm{AB},C_F}_{ij}$ corresponds
to the $C_F$
part of $\A\B_{ij}|_\mathrm{soft}$.
We again list our results in  six MATHEMATICA arrays, but each now having
3 dimensions. The first two indices  of an array run from 1 to 5 and
specifies the diagram. The last
index, n, labels the approximation;
the entry at n=1 is the substitution rule for one Feynman parameter as imposed by
the delta function for that diagram. One is however free to choose any Feynman
parameter except for the $\luv$ divergent diagrams where it has to be
$\b_2$ (see text).  Recall that in case of these diagrams the
substitution of $\b_2$ has already been
applied to $X_{\luv}$.  If, in a given limit,
a diagram does not diverge, the corresponding entry
in the array is zero. The arrays are defined in the following way:
\begin{align*}
\mathtt{CFaa[[ i, j, n ]]}=
\Big[\b\to 1-\mbox{$\sum$}\b,\: &X,  \:
X_\mathrm{cen},\: X_\mathrm{cen,soft},\:
X_\mathrm{soft},\: X_\mathrm{q},\:
X_\mathrm{q,soft},\\ 
&X_\mathrm{\bar
q}, \:  X_\mathrm{\bar q, soft},\: X_{\luv},\:
(X - X_\mathrm{soft})_\mathrm{is}
\Big]^{\mathrm{AA},C_F}_{ij} \\[.2cm]
\mathtt{ CAab[[ i, j, n ]]}=
\Big[\b\to 1-\mbox{$\sum$}\b,\: &X,\:
X_\mathrm{cen},\: X_\mathrm{cen,soft},\:
X_\mathrm{soft},\: X_\mathrm{q},\:
X_\mathrm{q,soft},\\
&X_\mathrm{\bar
q}, \:  X_\mathrm{\bar q, soft},\: X_{\luv},\:
(X - X_\mathrm{soft})_\mathrm{is}
\Big]^{\mathrm{AB},C_A}_{ij}\\[.2cm]    
\end{align*}
and, in complete analogy: 
$\:\mathtt{CFab[[ i, j, n ]]},\:\: \mathtt{CFbb[[ i, j, n ]]}\:$ and
$\:\mathtt{CAaa[[ i, j, n ]]},\:\:  \mathtt{CAbb[[ i, j, n ]]}$ .

The six arrays are available  in the files 
\begin{center}
 $\mathtt{CFaa.m},\:\: \mathtt{CFab.m},\:\:
\mathtt{CFbb.m}\quad$ and $\quad\mathtt{CAaa.m}, \:\:\mathtt{CAab.m},
 \:\:\mathtt{CAbb.m}$
\end{center}

The remaining expressions for the $\B\A$ diagrams are obtained by the 
replacements: 
\begin{align*}
\mathtt{CAba[[ i, j, n ]]} &= \mathtt{CAab[[ j, i, n ]]}\\
\mathtt{CFba[[ i, j, n ]]} &= \mathtt{CFab[[ j, i, n ]]}.
\end{align*}

{\bf C.\hspace{.5cm}}In order to calculate numerical values for the real 
corrections one has to perform the integrations over $\a, \al, \b_i$ 
which, as indicated in eqs.(\ref{eq:sum2CF}) and (\ref{eq:sum2CA}),
are interrelated by their limits of integration. For the numerical computation 
it is most suitable to have decoupled integrals; we therefore introduce new
variables $x_i$ such that the integrals extend from 0 to 1
\begin{align}
\label{eq:b-subst}
\al, \b_i \longrightarrow x_0, x_i\,.
\end{align}

For a given diagram the integrand is a sum of expressions,
($X-X_\mathrm{soft}-\ldots$), each of them being divergent in some
limit. In some cases this sum
has to be cast in an analytic form, suitable for the numerics.
We  list here those  integrands that we have used  for the  numerical integration.
For each diagram we provide the quantities $\Omega, \Omega_I$
such that the real corrections in 
(\ref{eq:sum2CF}) and (\ref{eq:sum2CA}) read:
\begin{align}
\label{eq:sum4CF}
& \left.  \Phi_{\gs}^{(1,{\mathrm real})}\right|_{C_F}^{\mathrm
 finite}= \frac{e^2e_f^2 \sqrt{N_c^2-1}  }{ (2 \pi)^6}
 \left(\frac{\pi^2}{24}\;(3 \pi^2 - 28) \right.\nonumber\\
&\left.\hspace{4cm} + \sum_{C_F}\: \int_0^1 d\a\; d x_0\: d x_1\: d
 x_2\: d x_3 \:\:\Omega^{C_F} \right)\,,  
\end{align}
\begin{align}
\label{eq:sum4CA}
& \left.  \Phi_{\gs}^{(1,{\mathrm real})}\right|_{C_A}^{\mathrm
 finite}= \frac{e^2e_f^2 \sqrt{N_c^2-1}  }{ (2 \pi)^6}
 \left(-\frac{\pi^2}{48}\;(3 \pi^2 - 28) \right.\nonumber\\
&\left.\hspace{4cm} + \sum_{C_A}\: \int_0^1 d\a\; d x_0\: d x_1\: d x_2\: d x_3\: 
  \left(\Omega\:- \Omega_I \right)^{C_A} \right)\,.  
\end{align}
The sums extend over the $C_F$ and $C_A$ parts of all diagrams.
$\Omega$ stands for the combination of $X$ and its subtractions, as 
required by eqs.(\ref{eq:sum2CF},\ref{eq:sum2CA}), expressed in terms of the 
new variables. $\Omega_I$ corresponds to the sixth line in
eq.(\ref{eq:sum2CA}); the substitution of $\al$ is different from that
in $\Omega$, as a result of the different limits in the $\al$
integration. Note that, in general, the $X$ depend on different $\b_i$, and 
the $\Omega$'s therefore are functions of different $x_i$'s,
as indicated in \eqref{eq:b-subst}. We list the $\Omega$'s as obtained from the
substitution \eqref{eq:b-subst} instead of renaming all variables into
$x_1,x_2,x_3$ as in eqs.(\ref{eq:sum4CF}, \ref{eq:sum4CA}). This allows
to localize the various limits discussed above.

The files
\begin{center}
$\mathtt{CAintegrand.f}$ \\
$\mathtt{CFintegrand.f}$ 
\end{center}

contain, as FORTRAN functions, the $\Omega$'s and $\Omega_I$'s for a minimal set 
of diagrams, separated into $C_A$ and $C_F$ parts.
All functions have, as arguments, $\br^2, s_0, \Lambda,
\a$ ($\mathtt{ R2,s0,LA,a}$), and four variables $x_i$. However, they do not always 
depend on all of them. 
The functions are labelled in accordance with the diagram they
correspond to. In case of lr-divergent diagrams we have combined
$\A\A_{ij}+\A\B_{ij}$ ($\B\B_{ij}+\B\A_{ij}$).

Labels of the $\Omega$'s are:
\begin{align*}
&\mathtt{Aabij}\::\quad  \A\B_{ij}^{C_A},
\hspace{1.3cm}\mathtt{Fabij}\::\quad  \A\B_{ij}^{C_F},\\
& \mbox{and correspondingly} \quad \mathtt{Aaaij}, \mathtt{Abaij}, \ldots,\\
&\mathtt{Aaabij}\::\quad  \A\A_{ij}^{C_A} + \A\B_{ij}^{C_A}, \hspace{1.3cm}
\mathtt{Abbaij}\::\quad  \B\B_{ij}^{C_A} + \B\A_{ij}^{C_A}.
\end{align*}

Labels of the $\Omega_I$'s are:
\begin{align*}
\mathtt{AaabijI}\::\quad  \A\A_{ij}^{C_A} + \A\B_{ij}^{C_A}, , \hspace{1.3cm}
\mathtt{AbbaijI}\::\quad  \B\B_{ij}^{C_A} + \B\A_{ij}^{C_A}.
\end{align*}

The function $\mathtt{Faa12(R2,LA,s0,a,x0,x1,x4,x7)}$, for instance,
that can be found in the file $\mathtt{CFintegrand.f}$ has to be inserted in
eq.\eqref{eq:sum4CF} as $\Omega^{C_F}$ in case of the
diagram $\A\A_{12}$.
All those functions which are not listed in the files $\mathtt{CAintegrand.f, CFintegrand.f}$
are either zero or can be obtained by exploiting the following symmetries
\begin{align*}
&\mathtt{Aaaij = Aaaji},\quad \mathtt{Aabij = Abaji},\quad
\mathtt{Abbij = Abbji}\,,\\
&\quad\mbox{and accordingly for the $C_F$ parts,} \\
&\mathtt{Aaabij + Abbaij = Aaabji + Abbaji},\\
&\mathtt{AaabijI + AbbaijI = AaabjiI + AbbajiI}.
\end{align*}


\end{document}